\documentclass[aps,prl,twocolumn,preprintnumbers,10pt,showpacs]{revtex4-1}
\usepackage{amsmath, amsthm, amssymb,bm,relsize}

\usepackage{graphicx}  
\usepackage{amssymb}   
\usepackage{amsmath}
\usepackage{hyperref}
\usepackage{xcolor}

\usepackage{endnotes}
\let\footnote=\endnote

\usepackage{graphicx}
\usepackage{psfrag}
\usepackage[caption=false]{subfig}
\usepackage{color,xcolor}
\definecolor{urlblue}{rgb}{0.2,0.4,0.7}
\definecolor{citegreen}{rgb}{0,0.6,0.2}
\definecolor{linkred}{rgb}{0.9,0.2,0.1}
\usepackage{hyperref}
\hypersetup{
colorlinks=true, citecolor=citegreen, linkcolor=blue,urlcolor=urlblue}

\usepackage{slashed}
\usepackage{array}
\newcolumntype{P}[1]{>{\centering\arraybackslash}p{#1}}

\allowdisplaybreaks

\def\zob{\overline{z}_1}
\def\ztb{\overline{z}_2}

\def\Aob{\overline A_1^I}
\def\Atb{\overline A_2^I}
\def\Athb{\overline A_3^I}

\def\Dobd{\overline D_{d,1}^I}

\def\Dtbd{\overline D_{d,2}^I}

\def\btob{\overline \beta_1}
\def\bttb{\overline \beta_2}

\def\lfr{\ln\left({\mu_F^2 \over \mu_R^2}\right)}
\def\lfrt{\ln^2\left({\mu_F^2 \over \mu_R^2}\right)}

\def\lqr{\ln\left({q^2 \over \mu_R^2}\right)}
\def\lqrt{\ln^2\left({q^2 \over \mu_R^2}\right)}

\def\lw{\ln(1-\omega)}
\def\w{\omega}
\def\one{\ln(\overline \omega)}
\def\two{\ln^2(\overline \omega)}
\def\three{\ln^3(\overline \omega)}
\def\four{\ln^4(\overline \omega)}

\def\aLqf{\ln\left({q^2 \over \mu_F^2}\right)} 
\def\aLqftwo{\ln^2\left({q^2 \over \mu_F^2}\right)}

\usepackage{endnotes}
\let\footnote=\endnote

\begin{document}

\preprint{IMSc/2018/05/03}
\preprint{DESY 18-067}

\title{Threshold resummation of the rapidity distribution for Drell-Yan production at NNLO+NNLL
}

\author{Pulak Banerjee$^{a,b}$}\email{bpulak@imsc.res.in}
\author{Goutam Das$^{c}$}\email{goutam.das@desy.de}
\author{Prasanna K. Dhani$^{a,b}$}\email{prasannakd@imsc.res.in}
\author{V. Ravindran$^{a,b}$}\email{ravindra@imsc.res.in}

\affiliation{
$^a$ The Institute of Mathematical Sciences, Taramani, Chennai 600113, India \\
$^{b}$ Homi Bhabha National Institute, Training School Complex, Anushakti Nagar, Mumbai 400085, India\\
$^c$ Theory Group, Deutsches Elektronen-Synchrotron (DESY), Notkestrasse 85, D-22607 Hamburg, Germany
}

\date{\today}

\begin{abstract}
We consider the production of pairs of lepton through the Drell-Yan process at the LHC and
present the most accurate prediction on their rapidity distribution.  While the fixed order 
prediction is already known to next-to-next-to-leading order in perturbative QCD,
the resummed contribution coming from threshold region of phase space up to next-to-next-to-leading logarithmic (NNLL) accuracy has been computed
in this article.  The formalism developed in~\cite{Catani:1989ne,Ravindran:2006bu,Banerjee:2017cfc}
has been used to resum large threshold logarithms in the two dimensional 
Mellin space to all orders in perturbation
theory.  We have done a detailed numerical comparison against other approaches that resum certain
threshold logarithms in Mellin-Fourier space.  
Our predictions at NNLL level are close to theirs 
even though at leading logarithmic and next-to-leading logarithmic level we differ.
We have also investigated the impact of these threshold logarithms on the stability of perturbation theory
against factorisation and renormalisation scales.  While the dependence on these scales does not get
better with resummed results, the convergence of the perturbative series shows a better trend compared
to the fixed order predictions.  This is evident from the reduction in the K-factor for the resummed case 
compared to fixed order. We also present the uncertainties on the predictions resulting from
parton distribution functions.
\end{abstract}

\maketitle

\textit{Introduction}.---
The Standard Model (SM) has been extremely successful in describing the physics of elementary particles.
The production of oppositely charged lepton-pairs, known as the 
Drell-Yan production~\cite{Drell:1970wh}, is one of the benchmark processes 
to probe physics at TeV energies at the colliders, namely earlier at Tevatron and now at the Large Hadron Collider (LHC).
Because of its large cross section and small systematic uncertainties, Drell-Yan production also serves as 
luminosity monitor~\cite{Khoze:2000db} at the LHC.
Most importantly, at hadron colliders, Drell-Yan production provides valuable information about the partonic structure of hadrons,
its clean electromagnetic probe is best suited for the search of any new physics beyond the SM (BSM). 
An excess rate over the SM in this channel will potentially indicate the signature of BSM physics. 
Drell-Yan is the potential background for processes involving $Z^\prime$ or $W^\prime$ and also for spin-2/graviton searches. 
With the current LHC-13 and upcoming LHC-14 runs, more events will be available to 
precisely study the Drell-Yan distributions over a wide kinematic region. 
%

Due to its undeniable importance, Drell-Yan has been studied theoretically to a great extent over many decades~\cite{Altarelli:1978id,Altarelli:1979ub}. There has been continuous efforts towards the computation of  higher order QCD and electroweak (EW) corrections to unprecedented accuracy. 
The full inclusive production cross section is known up to next-to-next-to-leading order (NNLO)~\cite{Hamberg:1990np,Harlander:2002wh} for a very long time. 
Very recently, predictions at next-to-next-to-next-to-leading-order (N$^3$LO) level have become available considering only the dominant soft-virtual (SV) contributions~\cite{Ahmed:2014cla,Li:2014bfa}. 
Electroweak corrections beyond leading-order (LO) are also known and at next-to-leading-order (NLO) level, they were computed in~\cite{Dittmaier:2001ay, Baur:2001ze}. 
While inclusive production is important for precise prediction of cross section, differential distributions 
allow a wider comparison with experiments. 
Fully differential distributions such as rapidity, transverse momentum for the Drell-Yan are known to up to NNLO level in QCD \cite{Anastasiou:2003yy,Anastasiou:2003ds,Catani:2009sm,Melnikov:2006kv,Gavin:2012sy}. 
SV contributions for the rapidity distribution are also known at N$^3$LO level 
\cite{Ravindran:2006bu,Ravindran:2007sv,Ahmed:2014uya}.
Studies where both QCD and EW corrections are combined can be found in \cite{Li:2012wna}.  
Parton showers matched with NLO QCD results for the Drell-Yan are also available in MC@NLO \cite{Frixione:2002ik}, POWHEG \cite{Frixione:2002ik,Frixione:2007vw} and
aMC@NLO \cite{Alwall:2014hca} frameworks.

One of the differential distributions that has been studied extensively is the transverse momentum ($p_T$) distribution of pair of leptons or the
vector bosons such as $Z/W^\pm$, see 
\cite{Ellis:1981hk,Arnold:1988dp, Gonsalves:1989ar,Mirkes:1992hu,Mirkes:1994dp}, often in their large $p_T$ region.
The rapidity distribution in Drell-Yan was computed in \cite{Ellis:1981hk} at NLO level in QCD and it was then extended to NNLO level
in  \cite{Anastasiou:2003yy,Melnikov:2006kv} which stabilise the predictions \cite{Anastasiou:2003ds} giving only a few percentage 
sensitivity to renormalisation and factorisation scales, say for example at the $Z$ mass region.  
However it has to be noted that the result does vary significantly w.r.t. the choices of different parton distribution functions (PDFs).  In particular, at large invariant mass or at large rapidity of
the final state, the cross sections are sensitive to large Bj\"orken $x$ regions of PDFs, where
different PDFs show not only differences between them but also exhibit large uncertainties.    
For a recent review see \cite{Accardi:2016ndt}.
This sensitivity of PDFs will in turn help to constraint the PDF sets much better.  Hence, it is important to study these distributions. Certain distributions in Drell-Yan production also helps to study unpolarized transverse momentum dependent PDFs. For the recent developments, see~\cite{Scimemi:2017etj,Scimemi:2018xaf}.
 

The fixed order predictions are often not reliable in certain regions of phase space where large logarithms of some kinematic variables can appear.
For example, at the partonic threshold {\it i.e.} where the initial partons have just enough energy to produce the final state such as a pair of leptons or 
$Z/W^\pm$ and soft gluons, 
the phase space available for the gluons become severely constrained which results in large logarithms.
These large logarithms however can be systematically resummed to all orders in perturbation theory for reliable predictions.
This has made the resummation program an important topic of investigation over many years.  For the inclusive production,  the resummation of
soft gluons in the threshold region was established 
\cite{Sterman:1986aj,Catani:1989ne,Catani:1990rp,Moch:2005ky,Laenen:2005uz,Idilbi:2005ni,Ravindran:2005vv} in the Mellin space
and for the transverse momentum distribution, at small $p_T$, the resulting large logarithms were shown to exponentiate in the
impact parameter space \cite{Collins:1984kg,Catani:2000vq}.  
A modern approach based on soft-collinear effective theory (SCET)
demonstrates similar resummation in momentum space, see \cite{Idilbi:2005ky} for inclusive production and \cite{Becher:2010tm} for transverse momentum distribution.
Resummation for the differential distribution with respect to the Feynman variable $x_F$ which describes the longitudinal momentum of the 
final state was studied in \cite{Catani:1989ne} and it was found that there were two thresholds and both could be resummed to all orders. For the resummation w.r.t. $x_F$ using a different scheme see~\cite{Westmark:2017uig}.  
For the rapidity distribution, resummation similar to the inclusive one, with a single scaling variable, can be obtained
in certain kinematic regions, see \cite{Laenen:1992ey, Sterman:2000pt, Mukherjee:2006uu,Bolzoni:2006ky,Bonvini:2014qga} 
and an equivalent approach based on SCET can be found in \cite{Becher:2006nr,Becher:2007ty}.
In the former one, called the standard direct QCD (dQCD) approach \cite{Catani:1989ne,Catani:1990rp,Sterman:1986aj},
since the resummation is performed in Mellin space where the phase space of the soft gluons factorises under appropriate Mellin transformation,
the threshold limit of the partonic scaling variable 
$z\to1$ corresponds to Mellin variable $N\to\infty$, where $z=q^2/\hat s$, $q^2=M_V, V=Z,W^\pm$ and $\hat s$ is the partonic centre of mass energy. 
In SCET approach \cite{Idilbi:2005ky,Becher:2006nr,Becher:2007ty}, however, resummation can be performed both in Mellin space as well as in $z$-space 
using the evolution operators of  soft  and the hard functions of the coefficient function. 


Resummation of large logarithms for rapidity distribution has been an interesting topic and several results are already available to very good accuracy. 
In \cite{Mukherjee:2006uu} the authors have studied resummation of rapidity $W^\pm$ in Mellin-Fourier (M-F) space following a conjecture (see \cite{Laenen:1992ey})
and later on same approach was used for Drell-Yan production in \cite{Bolzoni:2006ky}.
A more detailed study in the context of $W^\pm$ productions as well as production of a pair of leptons was undertaken in \cite{Bonvini:2010tp} emphasising the
role of prescriptions that take care of diverging series at a given logarithmic accuracy. 

In this article we will follow the dQCD approach \cite{Catani:1989ne}
to study soft gluon resummation for the 
rapidity distribution of a pair of leptons produced in hadron colliders.  Recently, using the formalism developed in \cite{Ravindran:2006bu,Ravindran:2007sv}, 
we \cite{Banerjee:2017cfc} derived a general result, applicable to production of any colorless state in hadron colliders, that resums the
soft gluons to all orders in perturbation theory in two dimensional Mellin (M-M) space spanned by $N_1,N_2$.   We also investigated their numerical 
impact on the rapidity distribution of the Higgs boson produced at the LHC.
The soft gluon effects show up through delta functions and plus distributions in the partonic cross sections 
when the partonic scaling variables reach the threshold limits, {\it i.e.}, $z_1\to 1$ and $z_2\to 1$
and these contributions
can be resummed to all orders both in $z_1,z_2$ space and in $N_1,N_2$ space.  
These resummed results were expanded to the desired accuracy to obtain fixed order predictions
for various observables~\cite{Ravindran:2007sv,Ahmed:2014uya,Ahmed:2014era} 
at the LHC in the SV approximation.
This double threshold limit, denoted by a pair of limits, namely ($z_1 \to 1$ , $z_2 \to 1$) corresponds to 
($N_1\to \infty$, $N_2\to \infty$) in M-M space.   The corresponding large logarithms are of the form $\ln^n (N_i)$, where $n=1,\cdot \cdot \cdot$ and $i=1,2$
and the resummation in M-M space
resums terms of the form $\w=a_s \beta_0 \ln(N_1 N_2)$ through a process independent function $g(\w)$ and a process dependent but 
$N_i$ independent function $g_0$. Here $\beta_0$ is the leading coefficient of the beta function of the strong coupling constant $g_s$
and $a_s = g_s^2(\mu_R^2)/16 \pi^2$ with $\mu_R$ being the renormalisation scale .  

The main goal of this article is to study the numerical impact of resummed
contributions in the M-M approach on the fixed order predictions
for the rapidity distribution of a pair of leptons in the Drell-Yan process 
at the LHC.  At NNLO level, fixed order results show remarkable stability against
the factorisation and renormalisation scales.  In addition,
they demonstrate excellent perturbative convergence.
While this is a good news for any phenomenological study with Drell-Yan process,
the question remains whether the fixed order predictions will be plagued
by presence of large kinematic logarithms resulting from soft gluons in the threshold regions 
at every order in perturbative expansion.   
The formalisms that can resum these large logarithms to all orders do exist and it
is not a priori clear whether the resulting resummed contributions will not affect 
the fixed order predictions.  Hence, a detailed study taking into account 
these threshold effects through resummation is warranted.  In addition,
owing to various ways by which these logarithms can be resummed, a detailed
comparison of these approaches is desirable.  This article attempts to address 
all these issues.  We start by recapitulating the resummation 
formalism based on M-M approach and then proceed with a detailed numerical 
study at the LHC and conclude with our findings. 

\textit{Theoretical framework}.--- 
In the QCD improved parton model, for the production of a pair of leptons with invariant mass $q^2$ and rapidity $y$, 
the double differential cross section can be written as  
\begin{eqnarray}\label{eq1}
{d^2 \sigma^q(\tau, q^2, y)\over dq^2 dy } &=&
\sigma^q_{\rm B}(x_1^0,x_2^0,q^2) 
\sum_{ab=q,\overline q}
\int_{x_1^0}^1 {dz_1 \over z_1}\int_{x_2^0}^1 {dz_2 \over z_2}~ 
\nonumber \\
&\!\!\hspace{-4.0cm}\times&
\hspace{-2.2cm}f_a\Big({x_1^0 \over z_1},\mu_F^2\Big)  f_b\Big({x_2^0\over z_2},\mu_F^2\Big)
\Delta^q_{d,ab} (z_1,z_2,q^2,\mu_F^2,\mu_R^2)\,,
\end{eqnarray} 
where $\sigma^q_{\rm B}(x_1^0,x_2^0,q^2) $ is the Born prefactor, $\tau=q^2/S=x_1^0 x_2^0$ with $q$ being the momentum of
the final state lepton pairs and $S=(p_1 +p_2)^2$ where $p_i$ are the momenta of the incoming hadrons.
The hadronic rapidity is defined as $y=\frac{1}{2}\ln \Big(\frac{p_2.q}{p_1.q}\Big)=\frac{1}{2}\ln \left(\frac{x_1^0}{x_2^0}\right)$;
$f_a\left({x_1^0 \over z_1},\mu_F^2\right)$ and   $f_b\left({x_2^0\over z_2},\mu_F^2\right)$ are the PDFs having momentum fractions $x_1 = {x_1^0/z_1}$, $x_2 = {x_2^0/z_2}$ respectively, renormalised at the factorisation scale $\mu_F$.
$\Delta^q_{d,ab} (a_s,z_1,z_2,q^2,\mu_F^2)$ (shorthand as $\Delta^q_{d,ab}(z_1,z_2)$) on the other hand is the Drell-Yan coefficient function for the rapidity distribution mass factorised at $\mu_F$. 
Unlike PDFs, these are calculable order by order in QCD perturbation theory in powers of $a_s$.
The coefficients in this power series expansion contain 
distributions such as $\delta(1-z_i)$ and  
$\left[ \frac{\ln^{m-1}(1-z_i)}{1-z_i}\right]_+$ with $m\leq 2n$, $n$ being the order of
perturbation and regular functions of $z_i$.  The former ones, namely the distributions, constitute the
SV part, denoted by $\Delta^{q,\text{SV}}_{d,ab}$ while the latter one, the  hard part is $\Delta^{q,H}_{d,ab}$.
In the SV part, these distributions result from certain regions of the phase space in the real emission sub processes 
and also of the loop integrals in the virtual ones.  While these distributions are singular as $z_i \rightarrow 1$,
they are integrable functions.  At the level of hadronic cross sections, they often  
dominate over the hard part when folded with the appropriate PDFs, in the above mentioned kinematic regions 
at every order in perturbation theory.
Hence, they can potentially disturb the reliability of the perturbative predictions.  The resolution is 
to resum these large terms, often the logarithms, to all orders to obtain any sensible prediction.
This is indeed the case for the rapidity distribution in Drell-Yan when the scaling variables
$z_1\to 1$ and $z_2\to 1$.  Recall that in the work by Catani and Trentadue \cite{Catani:1989ne}, a different distribution
namely Feynman $x_F$ for the Drell-Yan was studied in the context of threshold resummation and it was shown that
the potential threshold logarithms can be resummed to all orders in perturbation theory working in M-M space.
They had established that these logarithms could be exponentiated and also obtained the resummed result at the next-to-leading-logarithmic (NLL) accuracy.  Following this, in \cite{Ravindran:2006bu,Ravindran:2007sv,Ahmed:2014uya} 
we demonstrated for the rapidity distribution of
any colorless particle, resummation of the distributions defined with respect to 
the scaling variables $z_i$ to all orders in perturbation theory in $z_1,z_2$ space and 
later extended it to $N_1,N_2$ space in \cite{Banerjee:2017cfc} by applying two dimensional Mellin transformations on these distributions 
to obtain resummed result in the M-M space.  
The latter one, namely the resummed result in the M-M space, turned out to be more suitable for numerical
study and hence, we used this approach to demonstrate the importance of
these threshold logarithms for the rapidity distribution 
of the Higgs boson at the hadron collider to next-to-next-to-leading-logarithmic (NNLL) accuracy \cite{Banerjee:2017cfc} over NNLO.  In this paper, we explore this approach to study
the Drell-Yan process at the LHC.   We use the general result obtained for any colorless particle in \cite{Banerjee:2017cfc} 
to study the numerical impact on the rapidity distribution of a pair of leptons produced at the LHC.  
 
Note that the approach followed here \cite{Banerjee:2017cfc} differs from 
earlier ones (see \cite{Laenen:1992ey, Sterman:2000pt, Bolzoni:2006ky, Bonvini:2010tp, Becher:2006nr,Becher:2007ty}
in the way the threshold limit(s) is(are) taken). 
In the latter approach, the threshold contributions from soft gluons in the partonic cross section are 
defined by considering only those
distributions w.r.t. the scaling variable $z=z_1 z_2$ which appear in the region when $z \to 1$. 
The remaining contributions contain not only regular terms in $z$   
but also distributions and regular functions of  partonic rapidity variable ($y_p$).
Here, only distributions in $z$ are resummed to all orders treating the remaining terms as hard part. Thus the resummation for the Drell-Yan rapidity distributions  has been done using a single Mellin variable $N$ corresponding to $z$ and keeping
the $y_p$ dependent coefficients as it is.  
Interestingly, if one works in 
M-F space, it can be easily shown that in the limit $z\to 1$, 
the threshold logarithms resulting from $N\to \infty$ are identical to those of the inclusive cross section.
 Unlike the M-F approach where contributions resulting from $y_p\neq 0$ are dropped in the resummed formula, the methodology demonstrated in the present paper includes the threshold logarithms 
coming from $y_p\neq0$ region as well, hence covering wide range of values for the variable $y_p$.  The advantage of our approach is in constraining the PDFs at high momentum fraction.  In particular, di-lepton pair at large $y$ resulting from collisions in which one of the partons carries a large and the other a small momentum fraction $x$, can be used to constrain the PDFs at large $x$, a region not well constrained by the current results. 
 
We employ the technique developed in \cite{Banerjee:2017cfc} namely the M-M space approach 
to perform the soft gluon resummation for Drell-Yan rapidity distribution.   Thanks to the convolution structure of the hadronic cross section
in terms of the PDFs $f_{a,b}$ and the Drell-Yan coefficient functions $\Delta^q_{d,ab}$, the two-dimensional Mellin transformation of the
Born normalised hadronic cross section becomes  a simple product of $\tilde f_a(N_1)$, $\tilde f_b(N_2)$ and 
$\tilde \Delta^q_{d,ab}(N_1,N_2)$ where $\tilde f_c(N_i) = \int_0^1 dz_i z_i^{N_i-1}f_c(z_i)$ for $i=1,2, c=a,b$ and 
 \begin{eqnarray}\label{eq2}
 \hspace{-0.45cm}\tilde\Delta^q_{d,ab}(N_1,N_2) &= & \bigg[\prod_{i=1,2}    \int_0^1  dz_i   z_i^{N_i-1}   \bigg]\Delta^q_{d,ab} (z_1,z_2)
 \end{eqnarray}
It was shown in \cite{Ravindran:2006bu,Ravindran:2007sv,Banerjee:2017cfc} that the SV part of $\tilde \Delta^q_{d,ab}$ $(\tilde \Delta^{\text{SV}}_{d,q})$ exponentiates 
the threshold logarithms through the cusp anomalous dimension $A^q$ and the collinear functions $D^q_d$ giving the resummed result\\
\begin{widetext}
\begin{eqnarray}\label{eq3}
\tilde \Delta_{d,q}^{\text{SV}}(N_1,N_2) = g^q_{d,0}(a_s) \exp
	\Bigg(  
	\Big[\prod_{i=1,2} \int dz_i  &&~z_i^{N_i-1}   \Big] 
	\Bigg[~\delta(\ztb)~\Bigg({ 1 \over \zob} \Bigg\{ \int_{\mu_F^2}^{q^2 ~\zob}
{d \lambda^2 \over \lambda^2}~ A^q\left(a_s(\lambda^2)\right) 
	+ D^q_d\left(a_s(q^2~\zob)\right) \Bigg\}
\Bigg)_+  \nonumber\\&&
	+ {1 \over 2} \Bigg( {1 \over \zob \ztb } \Bigg\{A^q(a_s(q^2 \zob \ztb)) 
	+ {d D^{q}_d(a_s(q^2 \zob \ztb))\over d\ln (q^2 \zob \ztb)} \Bigg\}\Bigg)_+
+ (z_1 \leftrightarrow z_2)
\Bigg]
\Bigg)\,,
\end{eqnarray}
\end{widetext}
where $\overline z_i = (1-z_i)$.  
The cusp anomalous dimensions for the quark, $A^q$, 
are known for Drell-Yan up to 3-loops \cite{Korchemsky:1987wg,Korchemsky:1988hd,Korchemskaya:1992je,Kodaira:1981nh,Catani:1988vd,Moch:2004pa},
and the coefficients in $A^q = \sum_{i=1}^\infty a_s^i A^q_{i}$ are given by 

%
 \begin{eqnarray}
	 A^q_1 &=& 4 C_F \,, 
 \nonumber \\
	 A^q_2 &=& 8 C_F C_A \left( \frac{67}{18} - \zeta_2 \right) + 8 C_F n_f \left( -\frac{5}{9} \right) \,,
 \nonumber \\
	 A^q_3 &=& 16 C_F C_A^2 \left( \frac{245}{24} - \frac{67}{9} \zeta_2  + \frac{11}{6} \zeta_3
                              + \frac{11}{5} \zeta_2^2 \right)
 \nonumber\\ &&
             + 16 C_F^2 n_f \left( - \frac{55}{24} + 2 \zeta_3 \right)
             + 16 C_F C_A n_f 
 \nonumber\\ &&
	      \times \left( - \frac{209}{108} 
	     + \frac{10}{9} \zeta_2 - \frac{7}{3} \zeta_3 \right)
             - 16 C_F n_f^2 \left(  \frac{1}{27} \right) \,,
 \end{eqnarray}
and $D_d^q$s are related 
\cite{Ravindran:2006bu,Ravindran:2007sv,Banerjee:2017cfc} 
to the $D^q$s of the inclusive cross section for the Drell-Yan. Expanding $D^q_d$ as $D^q_d = \sum_{i=1}^\infty a_s^i D^q_{d,i}$, we find that
%
$		D^q_{d,1} = 0,
		D^q_{d,2} =
        C_f n_f   \big\{
           {112 \over 27}
          - {8 \over 3} \zeta_2
          \big\}
       + C_a C_f   \big\{
          - {808 \over 27}
          + 28 \zeta_3
          + {44 \over 3} \zeta_2
          \big\}$,
%
with the SU(N) color factors 
\begin{equation}
C_A=\text{N},\quad \quad \quad C_F= \frac{\text{N}^2-1}{2 \text{N}} , \quad \quad \quad
T_F= \frac{1}{2}
\end{equation}
and $n_f$ is the number of active flavours.
In \cite{Banerjee:2017cfc}, following \cite{Catani:2003zt}, we systematically computed the two dimensional Mellin transformations in the large $N_i$ limits 
and the result takes the following form: 
\begin{eqnarray}\label{eq5}
\label{eqn:expG}
 \tilde \Delta_{d,q}^{\text{SV}}(N_1,N_2)&=& \tilde g^q_{d,0}(a_s) \exp\Big( g_d^q (a_s,\w)\Big)\,,
\end{eqnarray}
where $\w$ is defined as $\w =  a_s \beta_0 \ln(\overline N_1 \overline N_2)$, with $\overline N_i = e^{\gamma_E} N_i, i=1,2$.  
The coefficients  $g_d (a_s,\w)$ are process independent and contain purely logarithmically enhanced terms and can be expanded as,
\begin{eqnarray}\label{eq6}
g^q_d(a_s,\w)=g^q_{d,1}(\w) \ln(\overline N_1 \overline N_2) + \sum_{i=0}^\infty a_s^i g^q_{d,i+2}(\w)\,.  
\end{eqnarray}
Rescaling the constants by appropriate $\beta_i$ as 
$\overline g^q_{d,1} = g^q_{d,1}$, $\overline g^q_{d,2} = g^q_{d,2}$, 
$\overline g^q_{d,3} = g^q_{d,3}/\beta_0$, 
$\overline A_i^q = A_i^q /\beta_0^i $, $\overline D_{d,i}^q = 
D_{d,i}^q/ \beta_0^i$ and $\overline \beta_i = \beta_i /\beta_0^{i+1}$, we find \cite{Banerjee:2017cfc}. 
\begin{widetext}
\begin{eqnarray}
\label{eqn:G}
\overline g^q_{d,1} &=& \Aob {1 \over \w} \Big\{\w +(1-\w) \lw\Big\} ,
\nonumber\\
\overline g^q_{d,2} &=& \w\Big\{  \Aob \btob -\Atb  \Big\}+ \ln(1-\w)\Big\{\Aob \btob + \Dobd - \Atb \Big\}+ {1\over2}\ln^2(1-\w) \Aob\btob    
	\nonumber\\&&
	+  \lqr\ln(1-\w)\Aob
	+ \lfr\w \Aob , 
\nonumber\\
\overline g^q_{d,3} &=&  - \frac{1}{2}\w\Athb - \frac{1}{2}\frac{\w}{1-\w} \bigg\{ -\Athb+(2+\w)\btob\Atb+\Big\{ (w-2)\bttb-\w\btob^2-2\zeta_2\Big\}\Aob+2\Dtbd-2\btob\Dobd \bigg\}
\nonumber\\&&
-\lw\bigg\{\frac{\btob}{1-\w}\Big\{ \Atb-\Dobd-\Aob\btob\w\Big\}-\Aob\bttb  \bigg\}+\frac{1}{2}\frac{\ln^2(1-\w)}{1-\w}\Aob\btob^2 +  \lfr \Atb\w
\nonumber\\&&
	- {1 \over 2}\lfrt \Aob  \w
 - \lqr\frac{1}{1-\w} \bigg\{\Big\{\Atb -\Dobd\Big\}\w-\Aob\btob\Big\{\w+\lw \Big\}\bigg\} 
	+{1 \over 2}\lqrt  \frac{\w }{1-\w} \Aob.
\end{eqnarray}
\end{widetext}
The first three coefficients of the QCD $\beta$ function, $\beta_0$, $\beta_1$ and $\beta_2$ are given by \cite{Tarasov:1980au} 
\begin{align}
\beta_0&={11 \over 3 } C_A - {4 \over 3 } T_F n_f \, ,
\nonumber \\[0.5ex]
\beta_1&={34 \over 3 } C_A^2-4 T_F n_f C_F -{20 \over 3} T_F n_f C_A \, ,
\nonumber \\[0.5ex]
\beta_2&={2857 \over 54} C_A^3 
          -{1415 \over 27} C_A^2 T_F n_f
          +{158 \over 27} C_A T_F^2 n_f^2
\nonumber\\[0.5ex]
&          +{44 \over 9} C_F T_F^2 n_f^2
          -{205 \over 9} C_F C_A T_F n_f
          +2 C_F^2 T_F n_f\,.  
\end{align}

The $N_1,N_2$ independent terms resulting from integrals have been absorbed in $\tilde g^q_{d,0}$
addition to $g^q_{d,0}$, which however depends on the specific process under study.
In principle, these $N_1,N_2$ independent terms can also be exponentiated.
For Drell-Yan production, expanding $\tilde{g}^{q}_{d,0} = \sum_{i=0}^{\infty} a_s^i \tilde{g}_{d,0}^{q(i)}$, we find up to $a_s^2$ order in strong coupling:
\begin{widetext}
\label{tgi}
	\begin{eqnarray}
		\tilde{g}_{d,0}^{q(0)} &=&
        1 \,,
\nonumber \\
		\tilde{g}_{d,0}^{q(1)} &=&
        C_F   \bigg\{
          - 16
          + 16 \zeta_2
          \bigg\}
       + \lfr C_F   \bigg\{
          - 6
          \bigg\}
       + \lqr C_F   \bigg\{
           6
          \bigg\}\,,
		\nonumber\\
		\tilde{g}_{d,0}^{q(2)} &=&
        C_F n_f   \left\{
		{127 \over 6}
		- {64 \over 3} \zeta_2
		+ {8 \over 9} \zeta_3
          \right\}
       + C_F^2   \bigg\{
		{511 \over 4}
		- 198 \zeta_2
	         - {60 \zeta_3}
		+ {552 \over 5} \zeta_2^2
          \bigg\}
       + C_A C_F   \bigg\{
		- {1535 \over 12}
		+ {376 \over 3} \zeta_2
		+ {604 \over 9} \zeta_3
		\nonumber\\&&
		- {92 \over 5} \zeta_2^2
          \bigg\}
       + \lfr\Bigg[ C_F n_f   \left\{
		{2 \over 3}
		+ {16 \over 3} \zeta_2
          \right\}
       +  C_F^2   \bigg\{
           93
           - 72 \zeta_2
          - 48 \zeta_3 
          \bigg\}
       +  C_A C_F   \bigg\{
		- {17 \over 3}
		- {88 \over 3} \zeta_2
          + 24 \zeta_3	
          \bigg\}\Bigg]
          \nonumber\\&&
       + \lfrt \Bigg[C_F n_f   \bigg\{
          - 2
          \bigg\}
       +  C_F^2   \bigg\{
           18
          \bigg\}
       + C_A C_F   \bigg\{
           11
          \bigg\}\Bigg]
       + \lqr \Bigg[C_F n_f   \bigg\{
		- {34 \over 3}
		+ {16 \over 3} \zeta_2
          \bigg\}
          \nonumber\\&&
       +  C_F^2   \bigg\{
          - 93
          + 72 \zeta_2
          + 48 \zeta_3
          \bigg\}
       +  C_A C_F   \bigg\{
	   {193 \over 3}
	   - {88 \over 3} \zeta_2
          - 24 \zeta_3  
          \bigg\}\Bigg]
       + \lqr \lfr C_F^2   \bigg\{
          - 36
          \bigg\}
          \nonumber\\&&
       + \lqrt \Bigg[C_F n_f   \bigg\{
           2
          \bigg\}
       +  C_F^2   \bigg\{
           18
          \bigg\}
       +  C_A C_F   \bigg\{
          - 11
	  \bigg\}\Bigg].
  \end{eqnarray}
\end{widetext}
To study the numerical impact of our resummed result, we require in addition 
fixed order results containing only the large logarithms to perform proper matching.  They 
can be obtained by truncating
the resummed result and in the following, we present the $\tilde \Delta^{\text{SV}(i)}_{d,q}$ by setting $\mu_R^2=\mu_F^2$ up to NNLO level:
\begin{widetext}
\begin{eqnarray}
	\tilde \Delta^{\text{SV}(0)}_{d,q} &=&
        1
         \,,
\nonumber\\
	\tilde \Delta^{\text{SV}(1)}_{d,q} &=&
        C_F 
  \bigg\{
          - 16
          + 16 \zeta_2
          \bigg\}
          + \two  C_F \bigg\{
           2
          \bigg\}
       +  \aLqf  C_F \bigg\{
           6
          \bigg\}
       +  \aLqf \one C_F  \bigg\{
          - 4
          \bigg\}
\,,
\nonumber\\
	\tilde \Delta^{\text{SV}(2)}_{d,q} &=&
        C_F n_f  
           \bigg\{
           \frac{127}{6}
           - \frac{64}{3} \zeta_2
          + \frac{8}{9} \zeta_3
          \bigg\}
          +C_F^2
          \bigg\{
           \frac{511}{4}
          - 198 \zeta_2
          - 60 \zeta_3
          + \frac{552}{5} \zeta_2^2
          \bigg\}
          +C_AC_F
          \bigg\{
          - \frac{1535}{12}
          + \frac{376}{3} \zeta_2
          + \frac{604}{9} \zeta_3
          \nonumber\\&&
          - \frac{92}{5} \zeta_2^2
          \bigg\}
          +  \one   \Bigg[
          C_F n_f\bigg\{- \frac{112}{27}\bigg\}
          +C_A C_F\bigg\{
           \frac{808}{27}
          - 28 \zeta_3
          \bigg\}
          \Bigg]
          +  \two   \Bigg[
         C_F n_f\bigg\{ - \frac{20}{9}\bigg\}
         +C_F^2 \bigg\{
          - 32
          + 32 \zeta_2
          \bigg\}
          \nonumber\\&&
            +C_A C_F \bigg\{
           \frac{134}{9}
          - 4 \zeta_2
          \bigg\}
          \Bigg]
          +  \three   \Bigg[
          C_F n_f\bigg\{- \frac{4}{9}\bigg\}
          +C_A C_F \bigg\{
           \frac{22}{9}
          \bigg\}
          \Bigg]
          +  \four  C_F^2 \bigg\{
           2
          \bigg\}
          +  \aLqf \Bigg[C_F n_f  \bigg\{
          -\frac{34}{3}
          \nonumber\\&&
          +\frac{16}{3}\zeta_2
          \bigg\}
          +C_F^2\bigg\{-93 + 72 \zeta_2 + 48 \zeta_3
          \bigg\} + C_A C_F\bigg\{\frac{193}{3}- \frac{88}{3} \zeta_2-24\zeta_3\bigg\}
          \Bigg]
          +  \aLqf \one   \Bigg[
          C_F n_f\bigg\{\frac{40}{9}\bigg\}
          \nonumber\\&&
          +C_F^2\bigg\{64
          - 64 \zeta_2\bigg\}
         +C_A C_F\bigg\{ - \frac{268}{9}
          + 8 \zeta_2\bigg\}
          \Bigg]
          +  \aLqf \two   \Bigg[
          C_F n_f\bigg\{ \frac{4}{3}\bigg\}
          +C_F^2\bigg\{12\bigg\}
          +C_A C_F\bigg\{-\frac{22}{3} \bigg\}
          \Bigg]
          \nonumber\\&&
          +  \aLqf \three 
          C_F^2  \bigg\{
          - 8
          \bigg\}
           +  \aLqftwo  \Bigg[C_F n_f \bigg\{ 2\bigg\}+ C_F^2 \bigg\{18 \bigg\}+C_A C_F\bigg\{ -11\bigg\} 
          \Bigg]
          \nonumber\\&&
          +  \aLqftwo 
       \one \Bigg[C_F n_f \bigg\{ -\frac{4}{3}\bigg\} + C_F^2 \bigg\{- 24\bigg\} + C_A C_F \bigg\{\frac{22}{3} \bigg\}
       \bigg]
          +  \aLqftwo \two  C_F^2 \bigg\{
           8
          \bigg\}.
\end{eqnarray}
\end{widetext}
where $\overline \omega = \overline N_1  \overline N_2$.  In the following, we will discuss how these resummed
contributions can be systematically included in order to study their phenomenological importance at the
LHC.

\begin{figure*}[htb]
    \centering
    \includegraphics[width=5in,height=3in]{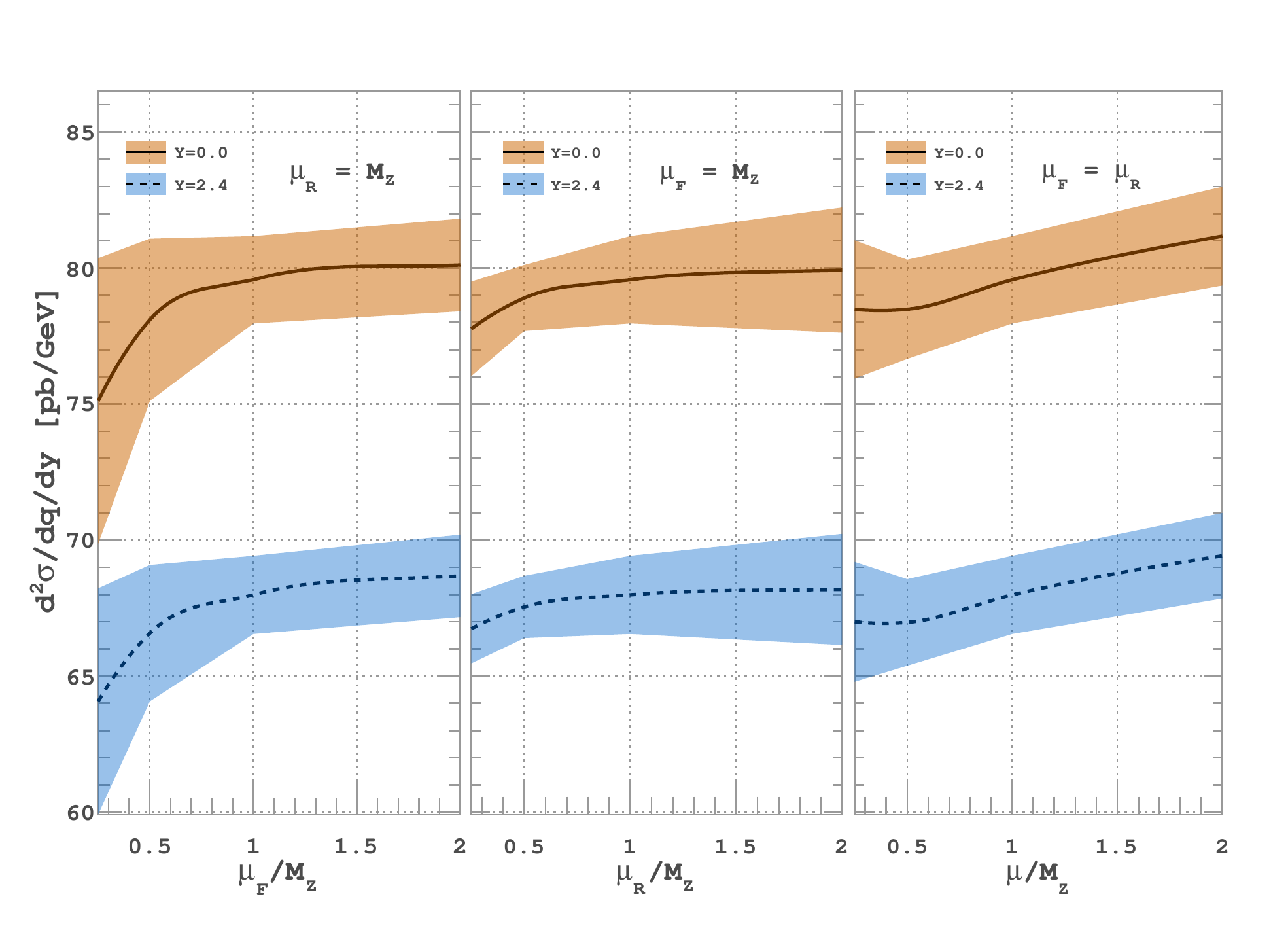}
\caption{Cross sections against $\mu_F$(left), $\mu_R$(middle) and $\mu$(right) variations at NNLO+NNLL 
for 14 TeV LHC.  The bands are obtained by using 7-point scale variation (see text for more details).}
    \label{fig1}
\end{figure*}

\begin{figure}[h!]
\centerline{
\includegraphics[width=8cm,height=6cm]{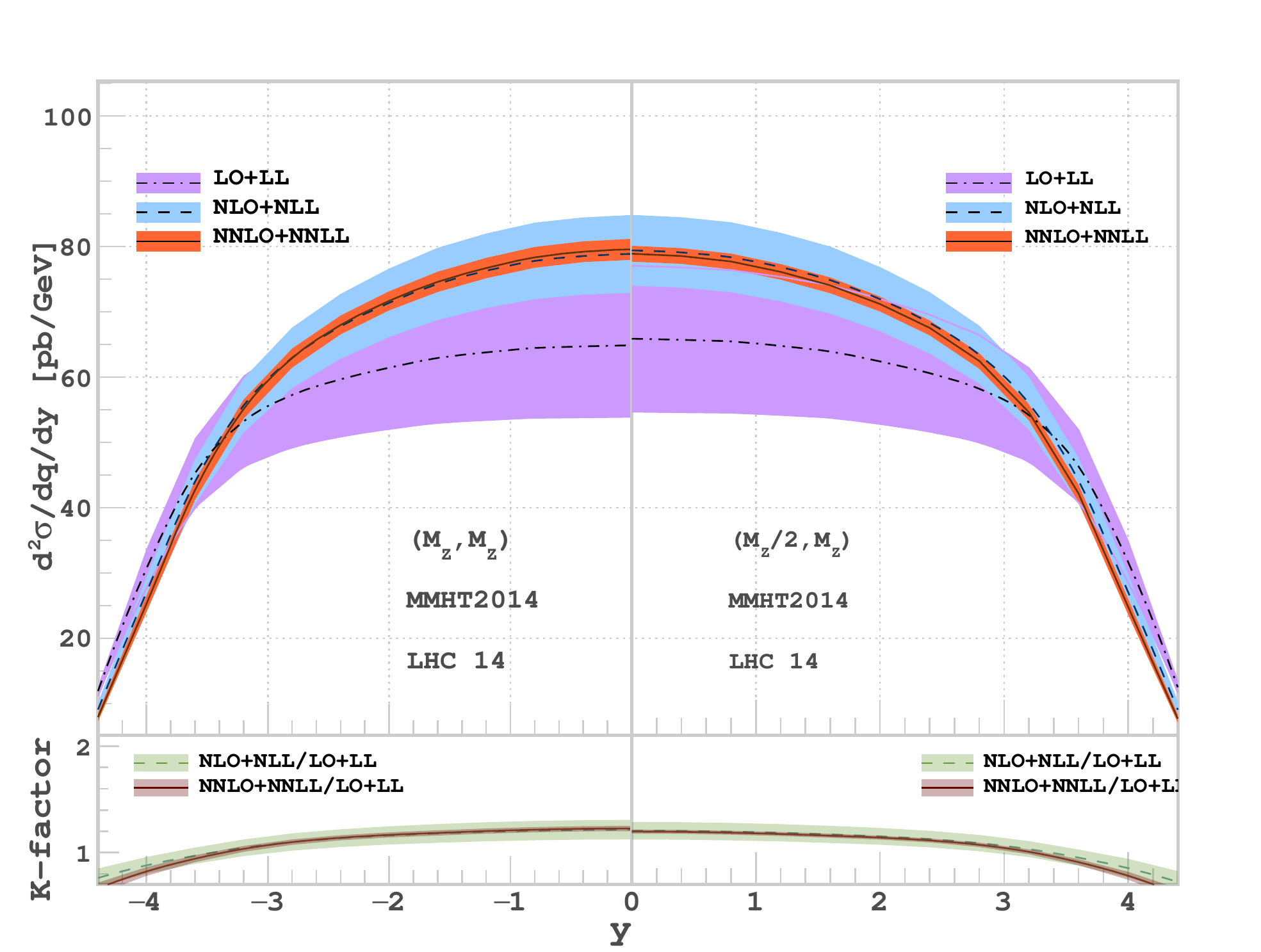}
}
\caption{Resummed rapidity distribution in Drell-Yan production for the two sets of central scale choices $(M_Z, M_Z)$ and $(M_Z/2, M_Z)$ using MMHT PDFs at 14 TeV LHC. Corresponding bands are obtained using 7-point scale variation around the central scale. The lower panel represents the corresponding K-factors.}
\label{fig9}
\end{figure}

\begin{figure}[b]
    \includegraphics[width=8cm,height=6cm]{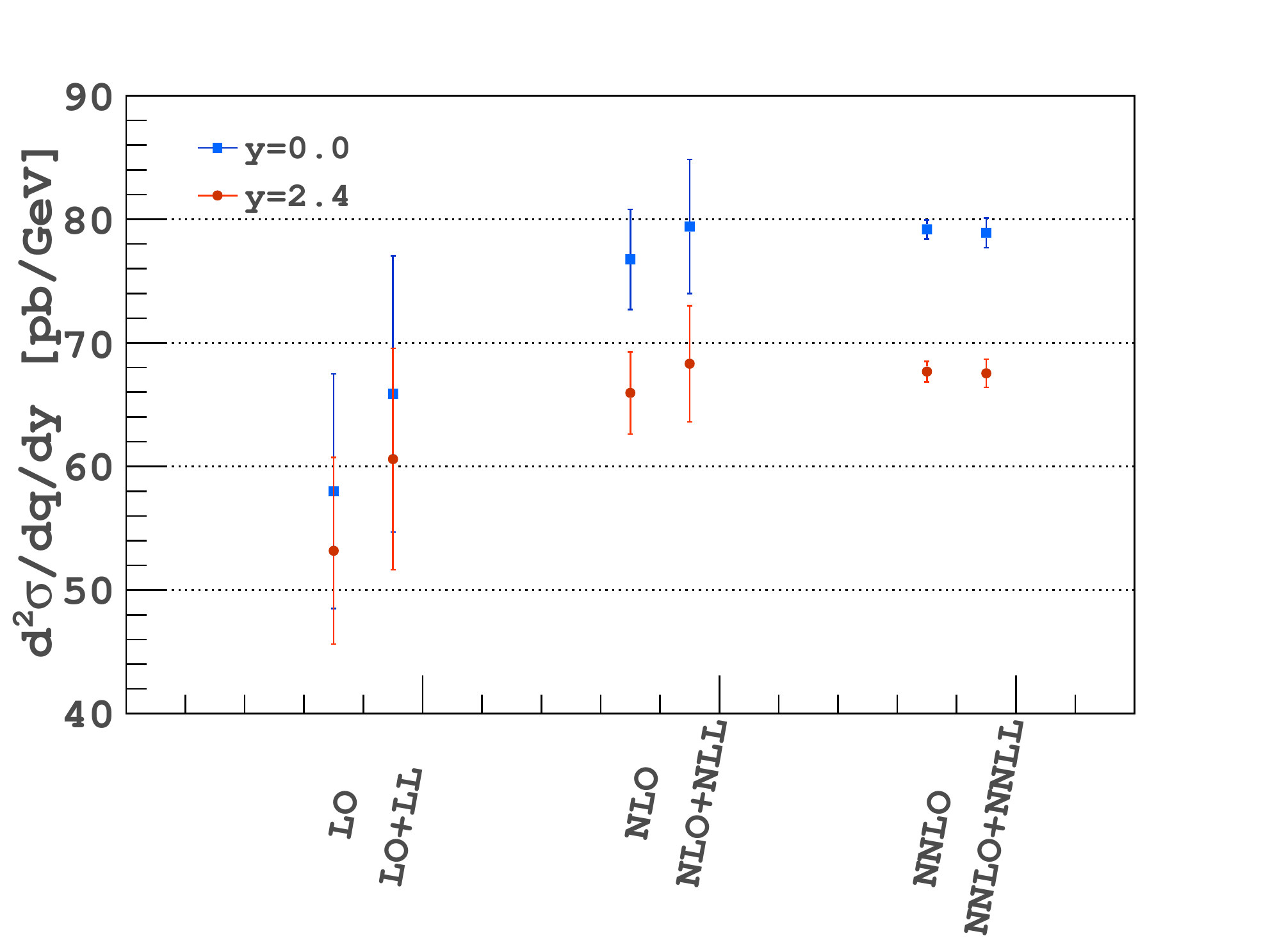}
\caption{Fixed order predictions with the central scale $\mu_R=\mu_F=M_Z$ and 
resummed prediction with the central scale $\mu_R=M_Z/2,\mu_F=M_Z$ for rapidity y=0 and 
y=2.4 using MMHT2014 PDF at each order. The uncertainties are obtained by using 7-point scale variation (see text for more details) around the central scale.}
    \label{fig2}
\end{figure}

\textit{Phenomenology}.---
Our next task is to include the resummed contributions 
consistently in the fixed order predictions and study their numerical impact on the rapidity distribution
of lepton pairs produced in the Drell-Yan process at the LHC. 
We consider the production of both leptons, {\it i.e.} $\ell^+\ell^-$, where $\ell=e,\mu$ 
through $Z$ and $\gamma^*$ in the collision of two hadrons at the centre of mass energy 14 TeV.
Unless otherwise stated, we will mostly focus on the region containing the $Z$-pole. 
We take $n_f = 5$ flavors, the MMHT2014(68cl) PDF set \cite{Harland-Lang:2014zoa} 
and the corresponding $a_s(M_Z)$ through the LHAPDF-6 \cite{Buckley:2014ana} interface 
at each order in perturbation theory.
For the fixed order rapidity distribution,
we use the publicly available code Vrap-0.9 \cite{Anastasiou:2003yy, vrap}.
The resummed contribution is obtained from $\tilde \Delta^{\text{SV}}_{d,q}(N_1,N_2)$ 
in Eq. (\ref{eq5}) after performing Mellin inversions which
are done using an in house Fortran based code.
Since the resummed result cannot be simply added to the fixed order one because 
all the $\ln(N_i)$ and  $N_i$ independent terms present in 
the resummed exponential $g^q_{d}$ and $\tilde{g}^q_{d,0}$ are already present
in the fixed order results and hence care is needed to avoid double counting.  
This can be achieved simply by employing a matching procedure
at every order.  The matched result is given below
\begin{widetext}
\begin{align}\label{eq8}
{d^2 \sigma^{q,\text{res}} \over dq^2 dy } =  
{d^2 \sigma^{q,\text{f.o}} \over dq^2 dy } +
\, {\sigma^{q}_B } \int_{c_{1} - i\infty}^{c_1 + i\infty} \frac{d N_{1}}{2\pi i}
 \int_{c_{2} - i\infty}^{c_2 + i\infty} \frac{d N_{2}}{2\pi i} 
e^{y(N_{2}-N_{1})}
\left(\sqrt{\tau}\right)^{2-N_{1}-N_{2}} 
\tilde f_{q}(N_{1}) 
 \tilde f_{q}(N_{2}) 
\Big[ \tilde \Delta_{d,q}^{\text{SV}} -  \tilde \Delta_{d,q}^{\text{SV}}\Big|_{\rm{trunc}} \Big] \,,
\end{align}
where $\sigma^q_B$ is given by
\begin{eqnarray}
\label{eq8a}
\sigma^{q}_B={4\pi \alpha^2 \over 3 q^4 N} \Bigg[e_q^2
-{2 q^2 (q^2-M_Z^2) e_q g_e^V g_q^V
\over
\left((q^2-M_Z^2)^2+M_Z^2 \Gamma_Z^2\right) c_w^2 s_w^2}
+{3 q^4 \Gamma_Z B^Z_l  \over  16 \alpha M_Z \left((q^2-M_Z^2)^2+M_Z^2 \Gamma_Z^2\right)
c_w^2 s_w^2 }\left(1+\Big(1-\frac{8}{3}s_w^2 \Big)^2 \right)\Bigg]
\end{eqnarray}
\end{widetext}
with $\alpha=\alpha(M_Z)=1/127.925$, $e_q$ is the quark charge, $M_Z = 91.1876$ GeV, $\Gamma_Z = 2.4952$ GeV, $s_w^2=0.227$, $c_w^2=1-s_w^2$, $g_e^V=-1/4+s_w^2$, 
$g_u^V=1/4-2/3s_w^2$, $g_d^V=-1/4+1/3s_w^2$, $B^Z_e = 0.03363$ and $B^Z_\mu=0.03366$. 
The first term in Eq.~(\ref{eq8}), 
$\left({d^2 \sigma^{q,\text{f.o}} / dq^2 dy }\right)$, 
corresponds to contributions 
resulting from a  fixed order perturbative computation.  
The second term on the other hand contains only threshold logarithms $\ln(N_i)$ 
but to all orders in perturbation theory. 
The subscript ``trunc" in the $\tilde \Delta_{d,q}^{\text{SV}}$ indicates that it is 
truncated at the same order as the fixed order one after expanding in powers of $a_s$.
Hence, at a given order $n$ in $a_s$ ({\it i.e.}, at order $a_s^n$), the non-zero 
contribution from the second term starts at order $a_s^{n+1}$ and includes terms 
from all orders.  For the fixed $n$-th order contribution, namely N$^n$LO, the contribution 
from the second term is called N$^n$LL . 
Hence, we use the notations LO, NLO and NNLO for the fixed order predictions and 
correspondingly LO+LL, NLO+NLL and NNLO+NNLL for the resummed ones.  
It is well known that the resummed expression diverges due to the missing 
non-perturbative contributions.  These divergences show up when $\omega \rightarrow 1$ in the 
functions $\overline g^q_{d,i}$; they are due to the coupling constant $a_s(\mu_R^2)$ that diverges
near the Landau pole.  In order to resolve this,
we have adopted the Minimal Prescription (MP) \cite{Catani:1996yz}.
The contours for the integrals corresponding to two Mellin inversions \cite{Anderle:2012rq} are chosen in such a way 
that all the poles in the complex plane 
spanned by $N_1,N_2$ remain to the left of the contours except for the Landau pole. 
Since the leading order contribution to the Drell-Yan process is due to EW interactions, the dominant 
theoretical uncertainty comes from the factorisation scale $\mu_F$ that enters through the parton distribution
functions while the dependence on the renormalisation scale $\mu_R$ starts only from NLO onwards. 
Unlike the leading order prediction, in the resummed case, the LL contributions do depend on
$\mu_R$ through $\omega$ in $\overline g^q_{d,1}(w)$ given in Eq.~(\ref{eqn:G}).  
Hence, $\mu_R$ dependence will show up at even LO+LL level. 
This will be evident from Fig.\ \ref{fig3} 
that one finds larger scale uncertainty from LO+LL contributions 
compared to the fixed order one at LO level.
It is then important to understand the impact of these two scales at the resummed level and also to determine
the optimal choice for the central scale around which the scale uncertainty remains minimal. 
For the fixed order case it has already been realised in \cite{Ebert:2017uel} that the optimal choice for the
central scale is when 
both $\mu_R$ and $\mu_F$ are set to $M_Z$.
In order to obtain the optimised central scale for the resummed case, 
we have plotted in Fig.~\ref{fig1} 
the dependence of the rapidity distribution on 
a) $(\mu_R = M_Z, \mu_F)$, b) $(\mu_R, \mu_F=M_Z)$
and finally c) $(\mu=\mu_R=\mu_F)$ at NNLO+NNLL level.
The symmetric band is obtained by performing  7-point scale variation \cite{Catani:2003zt,Bonvini:2010tp,Ebert:2017uel} 
around a given central scale with the constraint $(k_1,k_2)\otimes(\mu_R, \mu_F)_{\text{central}}$ where $(k_1,k_2)\in [1/2, 2]$ with $1/2 \le k_1/k_2\le 2 $ and
by taking maximum absolute deviation from the central scale.
From the first and the last panels of Fig.~\ref{fig1}, it is clear that the optimal central 
scale choice is $(M_Z,M_Z)$ whereas the middle panel favours $(M_Z/2,M_Z)$ for the central scale. 
Comparing all three panels, we find that 
the choice $(M_Z/2,M_Z)$ gives minimum uncertainty band.~However, to confirm the above analysis also holds true at each order in the perturbation theory, we have considered two different central scale choices $(M_Z,M_Z)$ and $(M_Z/2,M_Z)$ in Fig.~\ref{fig9}. 
We find that while the acceleration of the perturbative convergence are almost same for both cases,  uncertainty band at NLO+NLL and at NNLO+NNLL level are smaller for the central scale choice $(M_Z/2,M_Z)$ compared to the case $(M_Z,M_Z)$. In Fig.~\ref{fig2}, 
we compare predictions from the fixed order using the central scale $(M_Z,M_Z)$ against those 
from the resummed result using the central scale $(M_Z/2,M_Z)$ for two rapidities $y=0$ and $y=2.4$.
 The scale uncertainties from the resummed case at NNLO+NNLL are comparable to
what one obtains from  NNLO.  
However, the central values at NLO+NLL and NNLO+NNLL are very close to each other compared to those 
of fixed order results demonstrating the better perturbative convergence.
\begin{center}
\begin{table*}
\label{table1}
 \renewcommand{\arraystretch}{1.9}
\begin{tabular}{ |P{0.5cm} |P{1.4cm}||P{1.5cm}|P{1.5cm}|P{1.5cm}||P{1.5cm}|P{1.5cm}|P{1.5cm}||P{1.5cm}|P{1.5cm}|P{1.5cm}|}
 \hline
  y & $(\frac{\mu_R}{M_Z}, \frac{\mu_F}{M_Z})$ & $\text{LO}$ &$\text{LL}_{\text{M-F}}$ & $ \text{LL}_{\text{M-M}}$
    &$\text{NLO}$ & $\text{NLL}_{\text{M-F}}$ & $\text{NLL}_{\text{M-M}}$ & $\text{NNLO}$ & $\text{NNLL}_{\text{M-F}}$  & $\text{NNLL}_{\text{M-M}}$ \\
\hline
\hline
 0.0 & (2,\,2) & 72.626 &+0.988  & +3.219  & 73.450 &+1.639& +1.796 & 70.894 &+ 0.630 & +0.646  \\
 \hline
  0.0 & (2,\,1) & 63.197 &+0.768 & +2.595 & 70.625 &+0.761   &+1.017& 70.360 &+0.292 & +0.317\\
  \hline
 0.0 & (1,\,2) &  72.626 &+1.095 & +3.577 & 73.535 &+1.912 & +1.760  & 70.509 &+0.510 & +0.395 \\
  \hline
 0.0 & (1,\,1)  & 63.197 &+0.851 &+2.887& 71.395 &+0.858 & +0.901 & 70.537 &+0.248 & +0.167 \\
 \hline
 0.0 &(1,\,1/2) & 53.241 &+0.621 & +2.216 & 67.581 &+ 0.156&+0.140 & 69.834 &- 0.001 & - 0.094  \\
  \hline
  0.0 &(1/2,\,1)& 63.197 &+0.953 &+3.278& 72.355 &+0.945 &+0.681 & 70.266 &+0.091& - 0.015 \\
 \hline
 0.0 & (1/2,\,1/2) & 53.241 &+0.695 &+2.504& 69.259 &+0.102 &  - 0.154 & 70.283 &- 0.039& - 0.146 \\
 \hline
\end{tabular}
\caption{ Comparison of resummed results between M-F and M-M approach in the minimal prescription scheme at $y=0$ for various choices of scales.} \label{table1}
\end{table*}
\end{center}
As we have discussed in the introduction, in
\cite{Mukherjee:2006uu,Bolzoni:2006ky,Bonvini:2010tp} 
resummation of threshold logarithms for the rapidity distribution was achieved 
in the M-F space.  
Our formalism \cite{Banerjee:2017cfc} differs from the other approach 
in the way the threshold contributions are resummed. 
We resum large logs resulting from the regions 
where scaling variables $z_1$ and $z_2$ approach unity simultaneously while 
in the case of M-F, only large logarithms from the region where the partonic 
threshold variable $z$ approaches unity and the partonic rapidity $y_p$ is zero, are resummed.  
In the following we will make the numerical comparison of our predictions, namely the
M-M formalism against those of M-F reported in~\cite{Bonvini:2010tp}.
The fixed order contributions are obtained by using Vrap-0.9~\cite{Anastasiou:2003yy, vrap}; the resummed contributions up to NNLL 
for M-F are obtained by using publicly available code ReDY~\cite{redy} and for M-M, we use
our in house Fortran routine.  We have set 
all the parameters including the PDF set 
(NLO set of  NNPDF-2.0~\cite{Ball:2010de} at every order) 
same as those used in \cite{Bonvini:2010tp}. 
Both our results and those from ReDY are listed in the Table~\ref{table1} for various
scale choices at the central rapidity.   
At LL level, both M-F and M-M give positive contributions but the contribution from M-M 
is about three times larger compared to M-F independent of the scale choice. 
The additional contribution over LL at NLL for M-F is negative for some scale choices and positive for the
rest while for M-M, it is always negative.   The magnitude of these additional contributions 
for M-M is larger than M-F.  Interestingly, at NNLL level, the additional contributions over NLL
for M-F and M-M are both negative in a such a way that the net NNLL contributions from both approaches become 
comparable.  In the case of M-F, the NLO+NLL is 2\% larger compared to LO+LL and NNLO+NNLL is -4.7\% larger 
compared NLO+NLL.  For M-M, the corresponding ones are -0.8\% and -4.9\% respectively at $\mu_R=\mu_F=2 M_Z$.

In Fig.~\ref{fig3}, using Eq.~(\ref{eq8}), we present the cross section for producing lepton pairs as a 
function of the rapidity $y$ up to NNLO in the left panel and to NNLO+NNLL in the right panel 
along with the respective K-factors.  
The K-factor at a given perturbative order, say at N$^n$LO (N$^n$LO + N$^n$LL), is defined by the cross section at that order normalised by the same at LO (LO+LL) at the central scale $\mu_R=\mu_F=M_Z$.  
We have made this choice for the scales because the fixed order perturbative prediction is well behaved
around this scale \cite{Ebert:2017uel}.  The symmetric band 
at each order is obtained by varying 
$\mu_R$ and $\mu_{F}$ between $[M_{Z}/2, 2 M_{Z}]$ around the central scale $\mu_R=\mu_F=M_Z$ with the 
constraint $1/2\leq \mu_R/ \mu_F\leq 2$, 
by adding and subtracting to the central scale the highest possible uncertainties originating from all the scale combinations.
We find that the magnitude and the sign of the resummed contribution are sensitive to the order of perturbation 
as well the exact values of $y$ and the scales $\mu_R,\mu_F$.
For example, if we choose $\mu_R=M_Z/2$ and $\mu_F=M_Z$ instead of $\mu_R=\mu_F=M_Z$ as the central scale, 
we obtain a  negative contribution from NNLL terms for all values of rapidity. 

\begin{figure*}
\label{fig3}
\centering
\includegraphics[width=5in, height=3in]{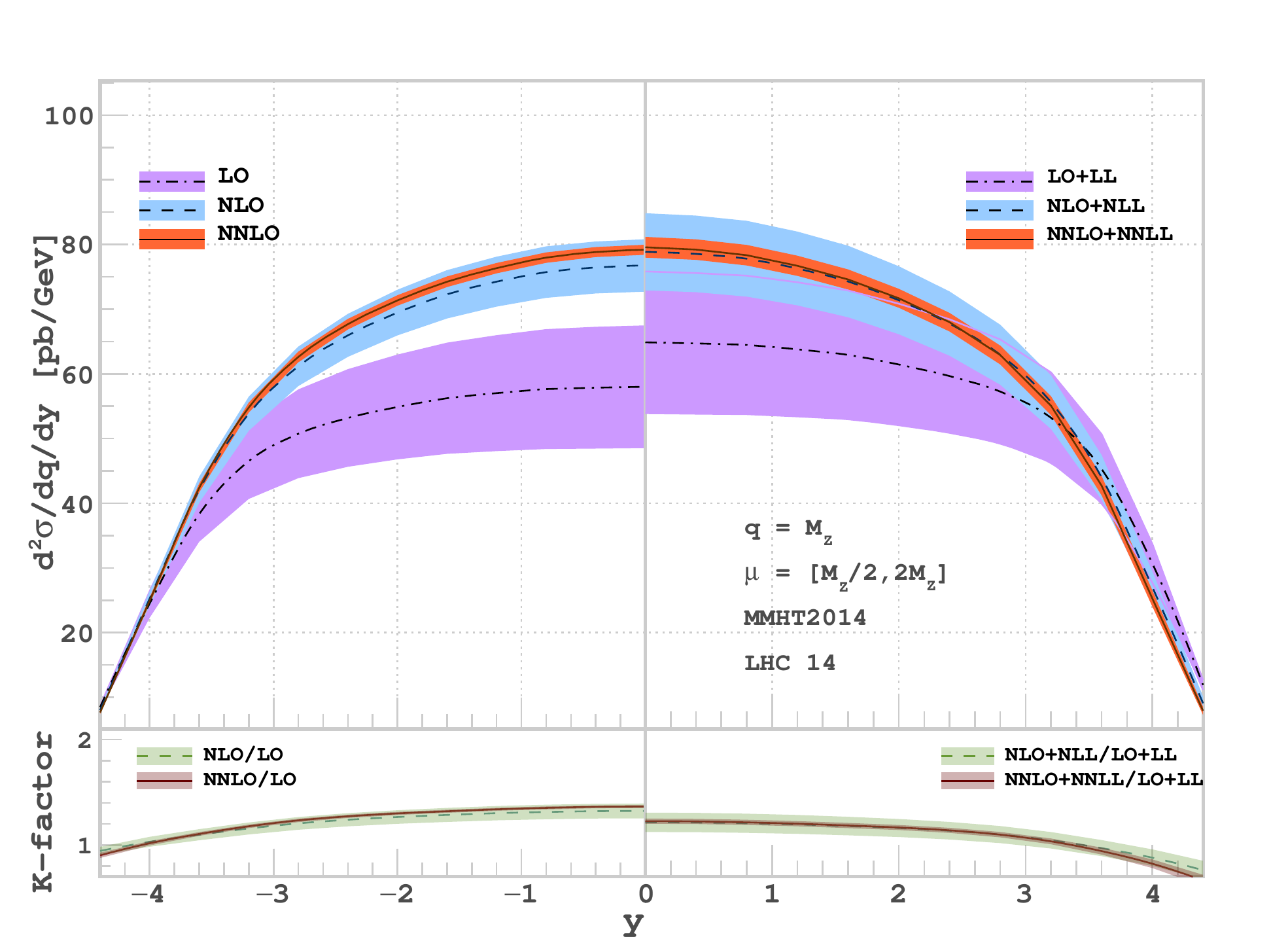}
\caption{Drell-Yan rapidity distribution for 14 TeV LHC at $q=M_Z$ using MMHT PDFs. 
The fixed order results are plotted in the left panel and the resummed results in the right panel. 
Central scale is chosen as $\mu_R=\mu_F=M_Z$ for both and the corresponding bands are obtained using 7-point scale variation (see text for more details) around the central scale.
The lower panel represents the corresponding K-factors.}\label{fig3}
\end{figure*}
\begin{figure*}
\label{figmur}
\centering
\includegraphics[width=5in, height=3in]{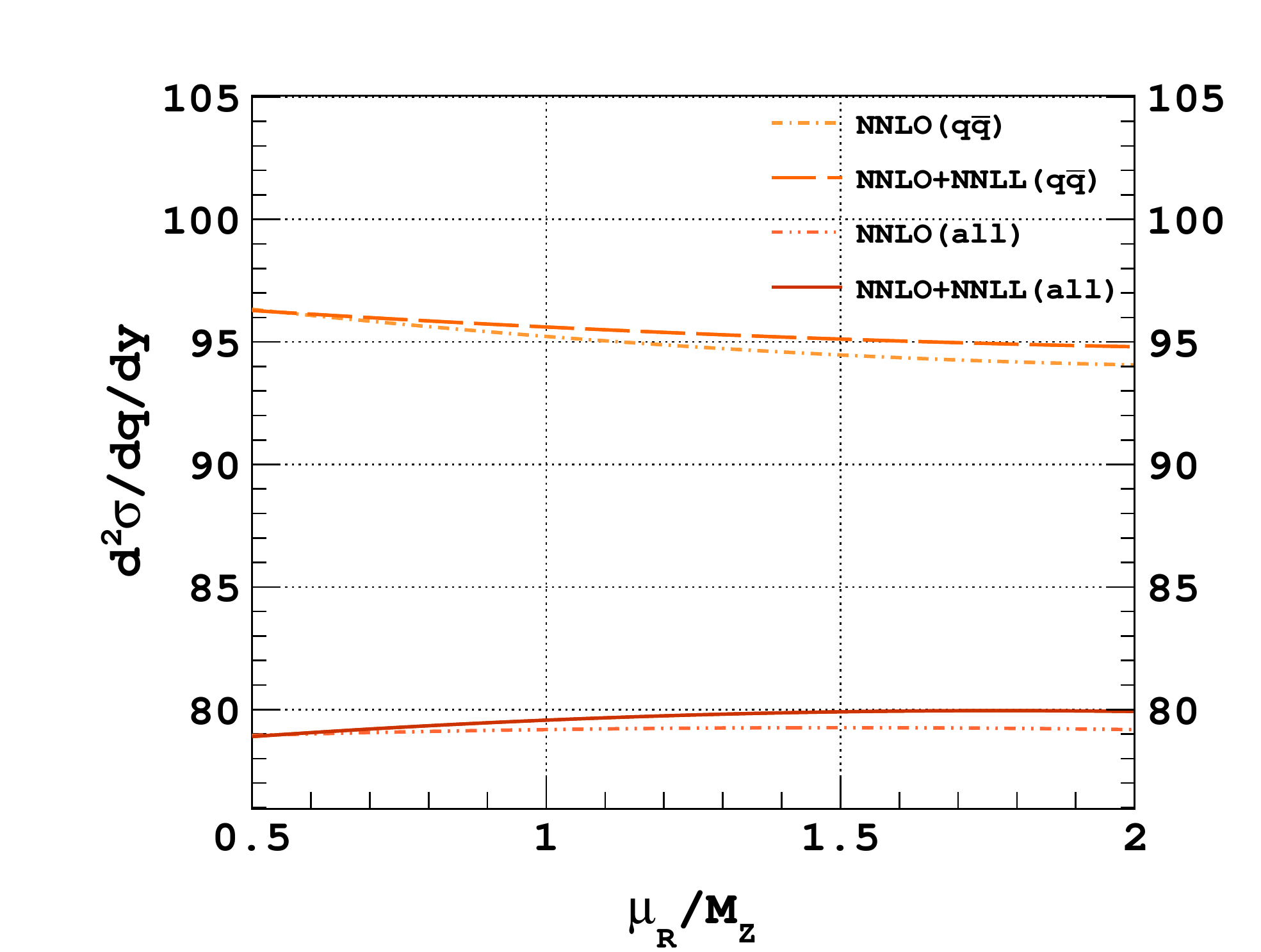}
\caption{Drell-Yan rapidity distribution for 14 TeV LHC at $y=0$ using MMHT PDFs. 
The variation of fixed order and resummed results as a function of $\mu_R$ are shown separately for $q\bar{q}$ channel and also for all the channels added together.}\label{figmur}
\end{figure*}

Fig.\ \ref{fig3} also
demonstrates that the inclusion of N${}^n$LL contributions increase the cross section at every order 
for a wide range of rapidity values.  In addition, 
the overlap among various orders is larger for the resummed case compared to the fixed order ones,   because the uncertainty band at each order in the resummed case is bigger compared to fixed order. 
As far as fixed order results are concerned, in particular at NNLO level, several partonic channels open up,
effectively reducing the scale uncertainty considerably.  On the other hand, resummed contributions 
come only from quark anti-quark initiated channels to all orders in perturbation theory 
as other channels do not give threshold logarithms of the type that is resummed.  
{We confirm this through Fig. \ref{figmur},
where we have studied the effects of resummation over the 
fixed order contributions, by considering  a) only  $q\bar{q}$ channel at NNLO and b) all the channels at NNLO. We perform our analysis for $y=0$ and set $\mu_F = M_Z$ while varying  $\mu_R$ between $M_Z/2$ to $2 M_Z$.
For the $q\bar{q}$ channel the resum contributions
arising from the two extreme scales are of opposite sign and their individual contributions are such that the NNLO+NNLL ($q\bar{q}$) curve shows a stable behaviour as compared to 
NNLO ($q\bar{q}$). While the fixed order decreases by 2.36\% from $M_Z/2$ to
$2 M_Z$, the corresponding decrease for NNLO+NNLL ($q\bar{q}$) is 1.53\%. This confirms the reduction of scale dependence upon adding resummed terms to the fixed order contributions.
To estimate the percentage corrections purely coming from the threshold region from this channel at each order of the perturbation theory, we have considered the case where the central scale is choosen to be $\mu_R = \mu_F = M_Z$.~As expected, at LO both fixed order and the truncated resummed predictions agree. But, at NLO and at NNLO we find truncated one is 7-8\% and 12-13\% larger compared to respective fixed order at the central rapidity region.~The largeness of the truncated results gets compensated by the -ve corrections coming from other channels emerging at respective orders.
However the scenario entirely reverses when we consider all the channels at NNLO. We find that the 
differential cross section at NNLO (all) increases by 0.29\% in the entire range of $\mu_R$ values;  the corresponding increase for NNLO+NNLL (all) is 1.29\%. This reduction of the scale dependence at NNLO is due to cancellations among different partonic channels.
However the resummation effects come only from $q\bar{q}$ channel which adds to the fixed order in such a way that the resummed uncertainty increases. This explains the increase of the scale uncertainty at each resummed order depicted in Fig. \ref{fig3}.}
Furthermore an
incomplete cancellation of the factorization scale dependence against the PDFs 
which do not contain resummed threshold logarithms also increases the band. For the threshold resummation effects in PDFs, see~\cite{Bonvini:2015ira}.
For the fixed order,  the K-factor at NLO varies between $1.3$ and $1.2$ and at NNLO between $1.37$ and $1.3$ 
over the entire rapidity region.  
On the other hand, the K-factors at both NLO+NLL and NNLO+NNLL significantly overlap with each other 
over most of the
regions of rapidity and stay around $1.2$.  This demonstrates a better perturbative convergence for resummed case
compared to the fixed order.  
In Table~\ref{table2}, 
\begin{widetext}
\begin{center}
\begin{table*}
\label{table2}
 \renewcommand{\arraystretch}{1.7}
\begin{tabular}{ |P{0.4cm}||P{1.85cm}|P{1.85cm}||P{1.7cm}|P{1.7cm}||P{1.7cm}|P{2.0cm}||P{0.7cm}|P{1.3cm}|P{0.9cm}|P{1.7cm}|}
 \hline
  y &LO&LO+LL&NLO&NLO+NLL&NNLO  & NNLO+NNLL&$\rm{K}_{\text{\tiny{NLO}}}$&$\rm{K}_{\text{\tiny{NLO+NLL}}}$&$\rm{K}_{\text{\tiny{NNLO}}}$&$\rm{K}_{\text{\tiny{\tiny{\tiny{NNLO+NNLL}}}}}$\\
 \hline
\hline
  0.0 &$58.002 {\scriptstyle \pm 16.36\%}$&$64.873{\scriptstyle \pm 16.89\%}$&$76.758{\scriptstyle\pm 5.28\%}$& $78.867{\scriptstyle\pm 7.56\%}$ &$79.182{\scriptstyle\pm 0.98\%}$   &$79.568{\scriptstyle\pm 2.02\%}$&1.323&1.216&1.365&1.226\\
  \hline
  0.8 & $57.645{\scriptstyle\pm 16.07\%}$  &$64.468{\scriptstyle\pm 16.61\%}$ &$75.727{\scriptstyle\pm 5.26\%}$&$77.797{\scriptstyle\pm 7.53\%}$&$77.968{\scriptstyle\pm 1.04\%}$ &$78.340{\scriptstyle\pm 2.03\%}$&1.314&1.207&1.352&1.215\\
  \hline
  1.6 &$56.228{\scriptstyle\pm 15.29\%}$&$62.929{\scriptstyle\pm 15.82\%} $&$72.295{\scriptstyle\pm 5.17\%}$&$74.274{\scriptstyle\pm 7.45\%}$&$74.239{\scriptstyle\pm 1.11\%}$ &$74.588{\scriptstyle\pm 2.08\%}$&1.286&1.180&1.320 &1.185\\
  \hline
  2.4& $53.181{\scriptstyle\pm 14.19\%}$&$59.655{\scriptstyle\pm 14.71\%}$ &$65.953{\scriptstyle\pm 5.04\%}$&$67.772{\scriptstyle\pm 7.33\%}$&$67.678{\scriptstyle\pm 1.21\%}$&$67.985{\scriptstyle\pm 2.11\%}$&1.240&1.136&1.273&1.140\\
\hline
\end{tabular}
\caption{Fixed order and the resummed cross sections with \% scale uncertainties along with
the K-factors at the central scale $\mu_R=\mu_F=M_Z$.} \label{table2}
\end{table*}
\end{center}
\end{widetext}
we have presented the cross section for benchmark rapidity values along with the 
percentage scale uncertainties.
Note that the differential cross-section at NNLO+NNLL level for the central scale is well 
approximated by the same at NLO+NLL. 
In fact, NNLO+NNLL increases approximately by 0.8\% with respect
 to NLO+NLL; the corresponding number for NNLO over NLO is approximately 3\%.
From the trend that resummed results give, 
we anticipate N$^3$LO+N$^3$LL cross-section will fall completely within the NNLO+NNLL band.

In Fig.~\ref{fig4},
\begin{figure*}
\label{fig4}
\includegraphics[width=5in, height=3in]{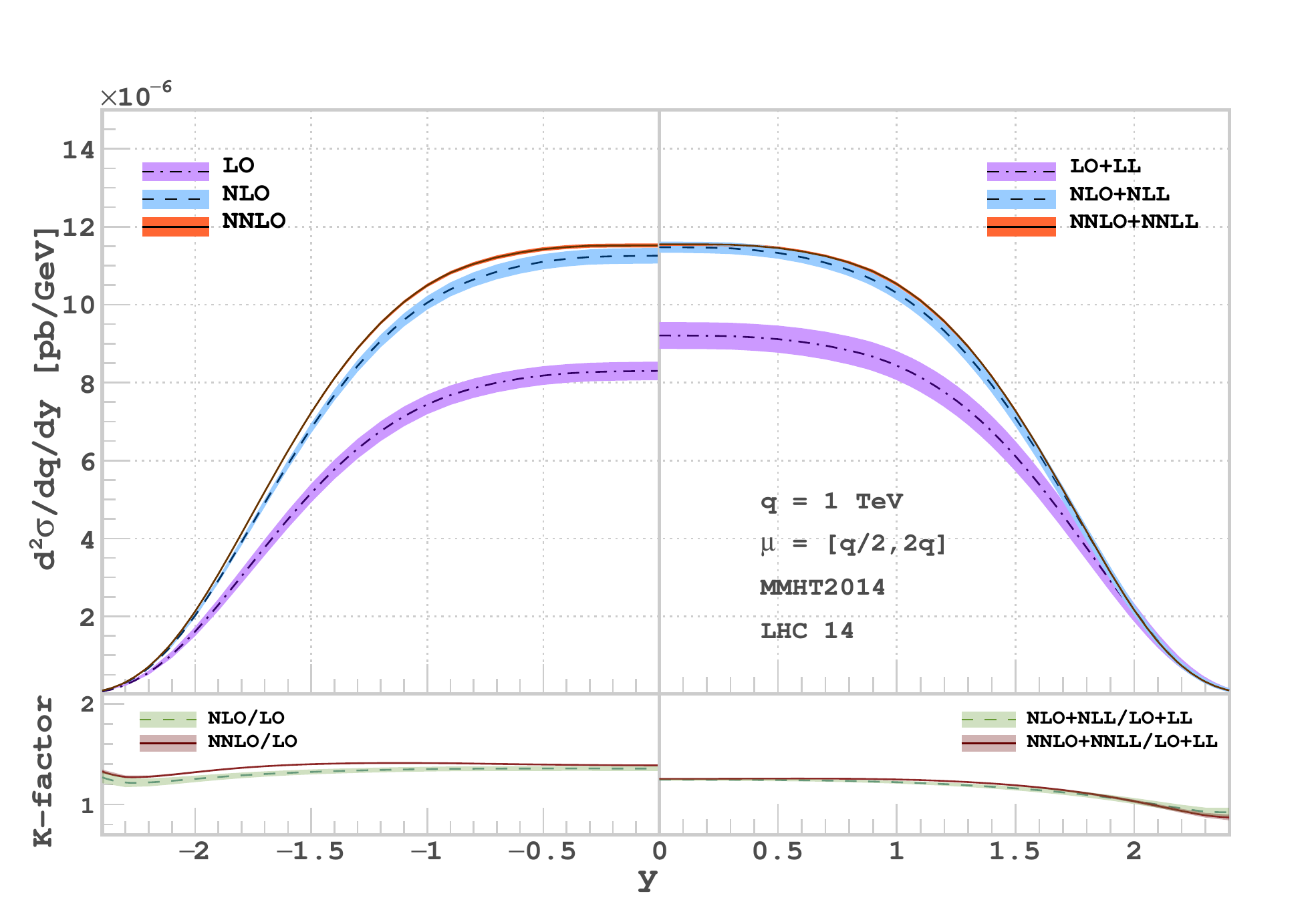}
\caption{Same as Fig.\ \ref{fig3} but for $q=1$ TeV.}\label{fig4}
\end{figure*}
we have plotted both fixed order and resummed results at various orders for the larger invariant mass at $q=1$ TeV.
Interestingly, the uncertainty bands at NLO+NLL and NNLO+NNLL levels are better compared to those from fixed order. 
Also the predictions at various orders are closer compared to those from fixed order demonstrating better
perturbative convergence of the higher order predictions from the resummed terms.
In fact the resummed K-factor for the central rapidity at NNLO+NNLL is 1.25 compared to 1.39 at NNLO.

\begin{figure}
\label{fig5}
\includegraphics[width=9cm, height=6cm]{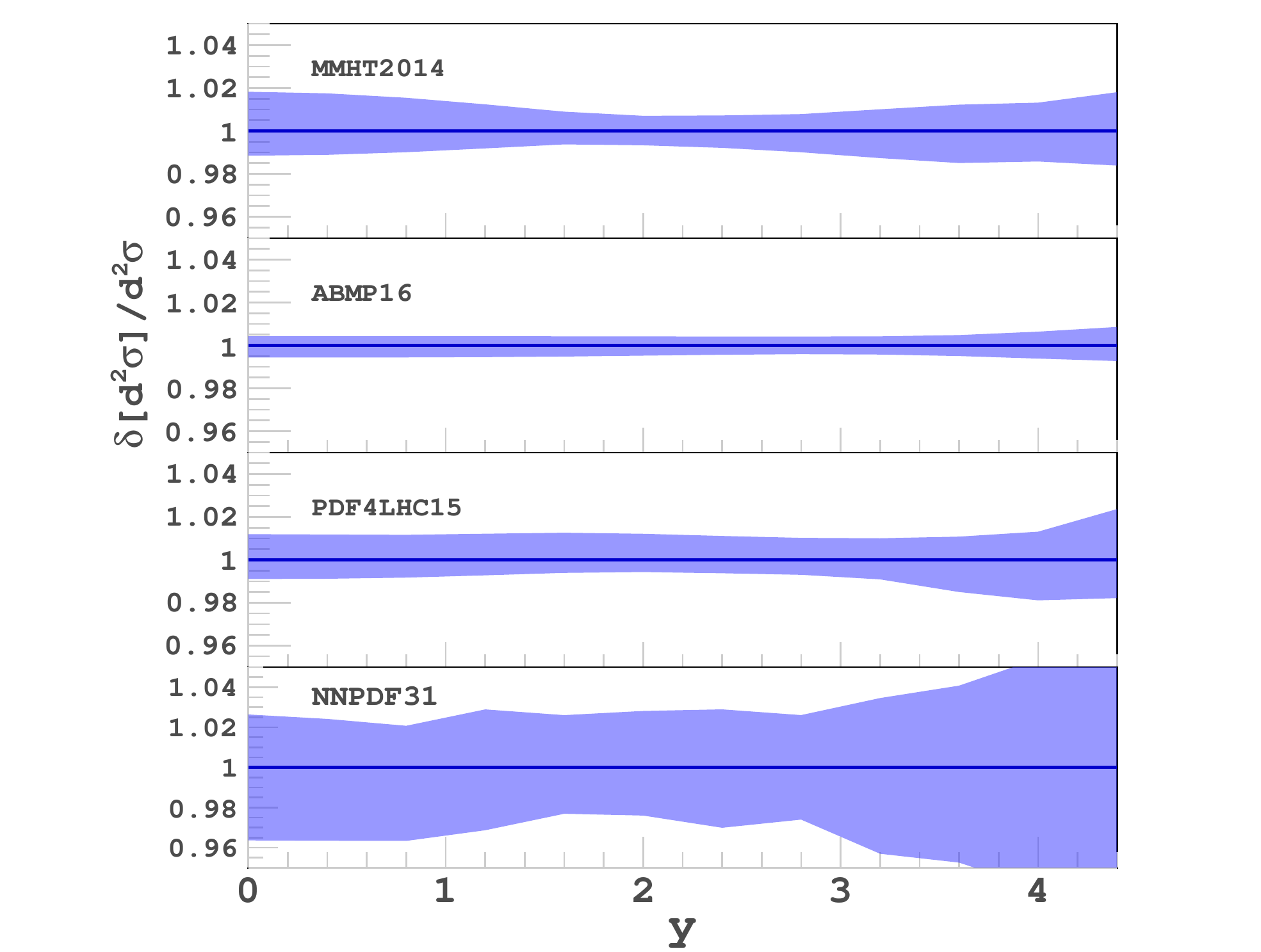}
\caption{PDF variation at NNLO+NNLL using various sets. The y-axis represents the ratio of extremum variation over the central PDF set.}\label{fig:pdfvar}
\end{figure}

As there are several PDF groups in the literature, each providing sets of PDFs,
it is customary to estimate the uncertainty resulting from the choice of PDFs within 
each set of a given PDF group.  Using PDFs from different PDF groups 
namely MMHT2014nnlo68cl \cite{Harland-Lang:2014zoa}, 
ABMP \cite{Alekhin:2017kpj}, NNPDF3.1 \cite{Ball:2017nwa} and PDF4LHC\cite{Butterworth:2015oua}  
we have obtained the cross sections along with the corresponding PDF uncertainty.  
In Fig.\ \ref{fig:pdfvar}, we have plotted the uncertainty bands for various PDF sets
as function of rapidity in order to demonstrate the correlation of PDF uncertainty with the rapidity values.  
This will help to better constrain the PDF fits using measurements on rapidity in the Drell-Yan process.
In Table \ref{table3}, we have also tabulated the cross sections along with \% uncertainties resulting from the choice of different 
PDFs.  

\begin{table}
\label{table3}
\renewcommand{\arraystretch}{1.7}
\begin{tabular}{ |P{0.5cm}||P{1.8cm}|P{1.8cm}|P{1.8cm}|P{1.8cm}| }
 \hline
  y &MMHT  & ABMP  &NNPDF & PDF4LHC  \\
 \hline
\hline
  0.0 &  $79.568^{+1.83\%}_{-1.16\%}$   &$79.756^{+0.43\%}_{-0.56\%}$ & $81.959^{+2.64\%}_{-3.64\%}$ &$78.734^{+1.20\%}_{-0.89\%}$  \\
  \hline
  0.8 &  $78.340^{+1.55\%}_{-0.99\%}$  & $78.202^{+0.43\%}_{-0.56\%}$ & $80.256^{+2.07\%} _{-3.66\%}$&$77.390^{+1.17\%}_{-0.83\%}$ \\
  \hline
  1.6 &$74.588^{+0.90\%}_{-0.63\%}$ &$73.738^{+0.42\%}_{-0.52\%}$  &$75.178 ^{+2.61\%}_{-2.31\%}$&$73.505^{+1.26\%}_{-0.61\%}$  \\
  \hline
  2.4&$67.985^{+0.72\%}_{-0.79\%}$ &$66.653^{+0.41\%}_{-0.44\%} $ & $67.354^{+2.89\%}_{-3.01\%}$ &$67.070^{+1.11\%}_{-0.62\%} $\\
\hline
\end{tabular}
\caption{Cross sections at NNLO+NNLL using different PDF sets along with percentage uncertainties 
for  $y=0, 0.8, 1.6, 2.4$.}
\label{table3}
\end{table}

\begin{figure}
\label{fig6}
\includegraphics[width=8cm, height=5cm]{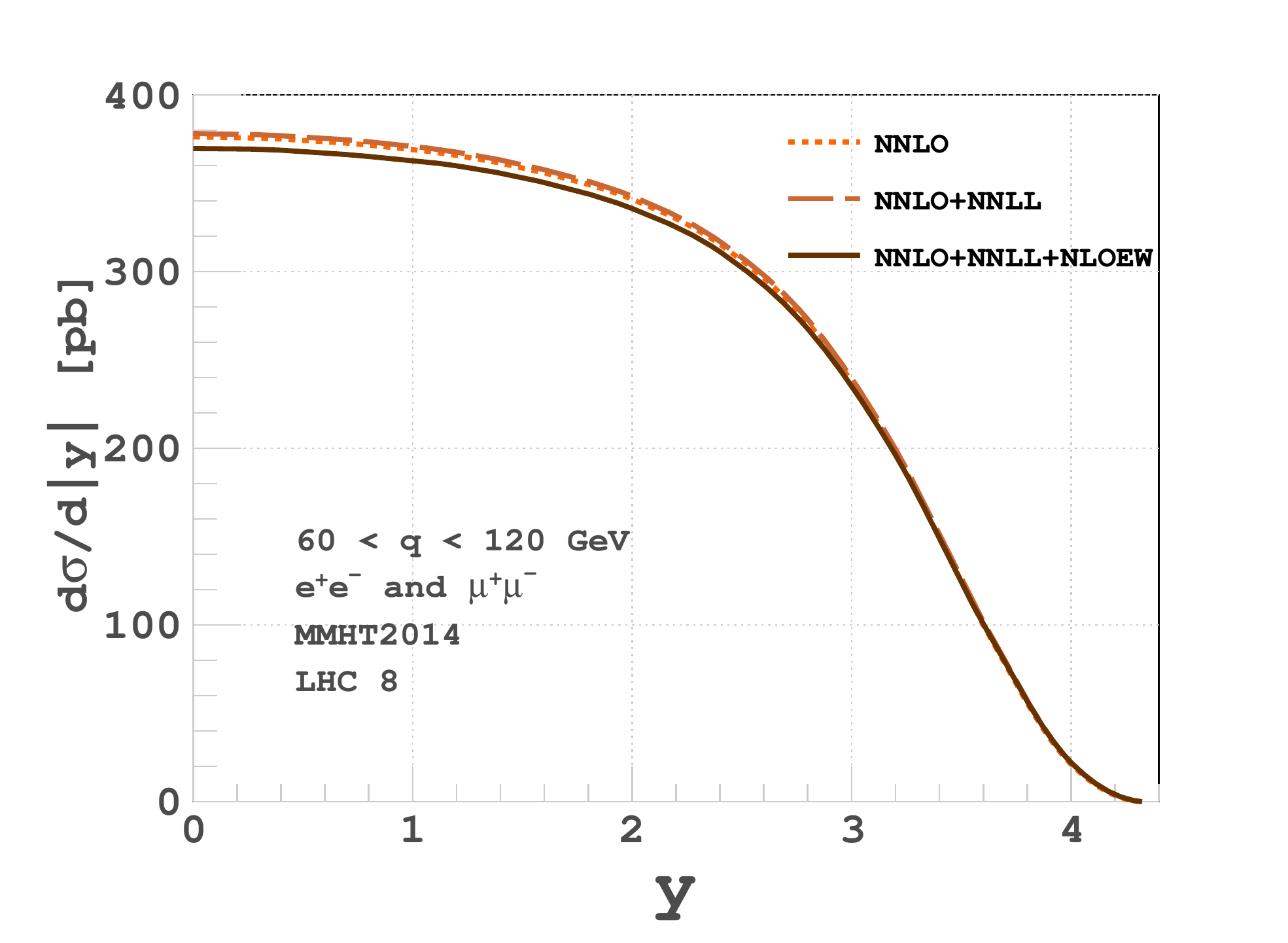}
\caption{Rapidity distribution at NNLO+NNLL for 8 TeV LHC in the invariant mass range $60<q<120$ GeV.
The dotted line is the fixed order NNLO contribution, the dashed line represents NNLO+NNLL result and the solid line includes EW corrections.
}\label{fig:qint}
\end{figure}

We have studied the $q$-integrated rapidity distribution 
at the LHC with 8 TeV centre of mass energy at NNLO+NNLL.
The invariant mass is integrated between $60$ GeV and $120$ GeV and choose $\mu_R=\mu_F=M_Z$. 
Unlike our earlier analysis, we have included both 
$e^+e^-$ as well as $\mu^+\mu^-$ final states. 
NLO EW corrections
are also included as they are also comparable to QCD corrections at NNLO+NNLL.
The EW contributions are obtained by using the publicly available code 
Horace-3.2~\cite{horace, Alioli:2016fum, CarloniCalame:2007cd, CarloniCalame:2005vc}. 
We use the $G_{\mu}$ scheme and take $G_F=1.16639\times 10^{-5}$, 
$M_W=80.395$ GeV, $M_Z=91.1876$ GeV and 
use MMHT2014nlo68cl pdf. 
The electron and muon masses are taken 
to be $m_e = 0.51099$ MeV and $m_{\mu} =0.10566$ GeV respectively. 
The NNLL contribution increases the cross-section  
by roughly around 0.5\% with respect to NNLO. 
However the EW corrections at NLO give 
negative contribution to the cross-section. 
The corrections are different for $e^+e^-$ and $\mu^+\mu^-$ pairs. 
For electrons, the EW contributions are twice that of muons.
In total from the electron and muon channels, 
we see an overall 2.3\%  decrease in the cross section w.r.t. the NNLO in the central rapidity region. 
%
The rapidity distribution in Fig.\ \ref{fig:qint} being inclusive in transverse momenta 
of the final state leptons, can not be directly compared 
with the results presented in \cite{CMS:2014jea} where a minimum transverse momenta 
cut is applied in the selection of final state leptons. 
To really compare this one needs distributions exclusive of transverse momenta which at the moment beyond the scope of the current paper and we leave it to future work.
\begin{widetext}
\begin{center}
\begin{figure}[h!]
\includegraphics[height=6cm,width=8cm]{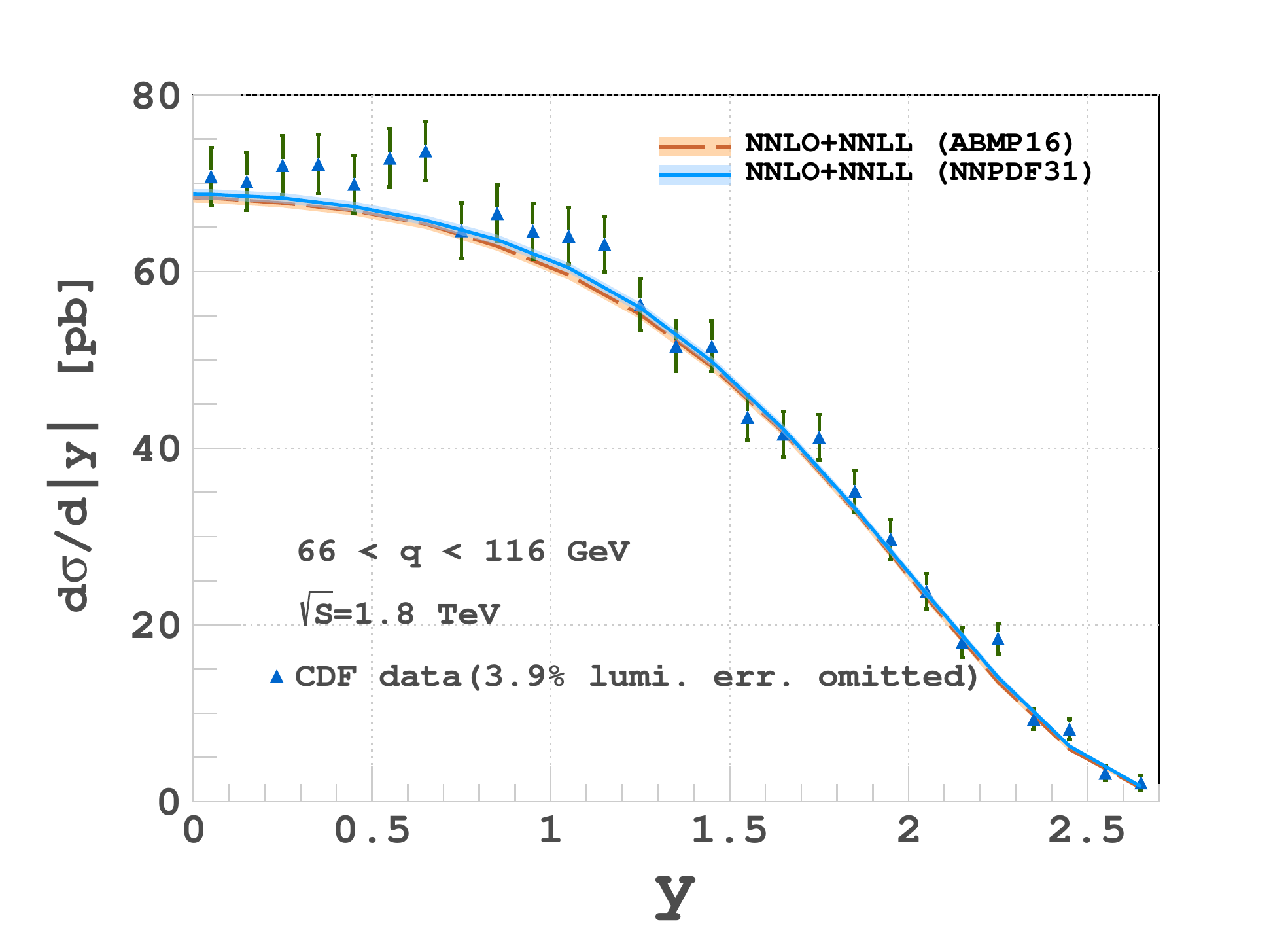} 
\includegraphics[height=6cm,width=8cm]{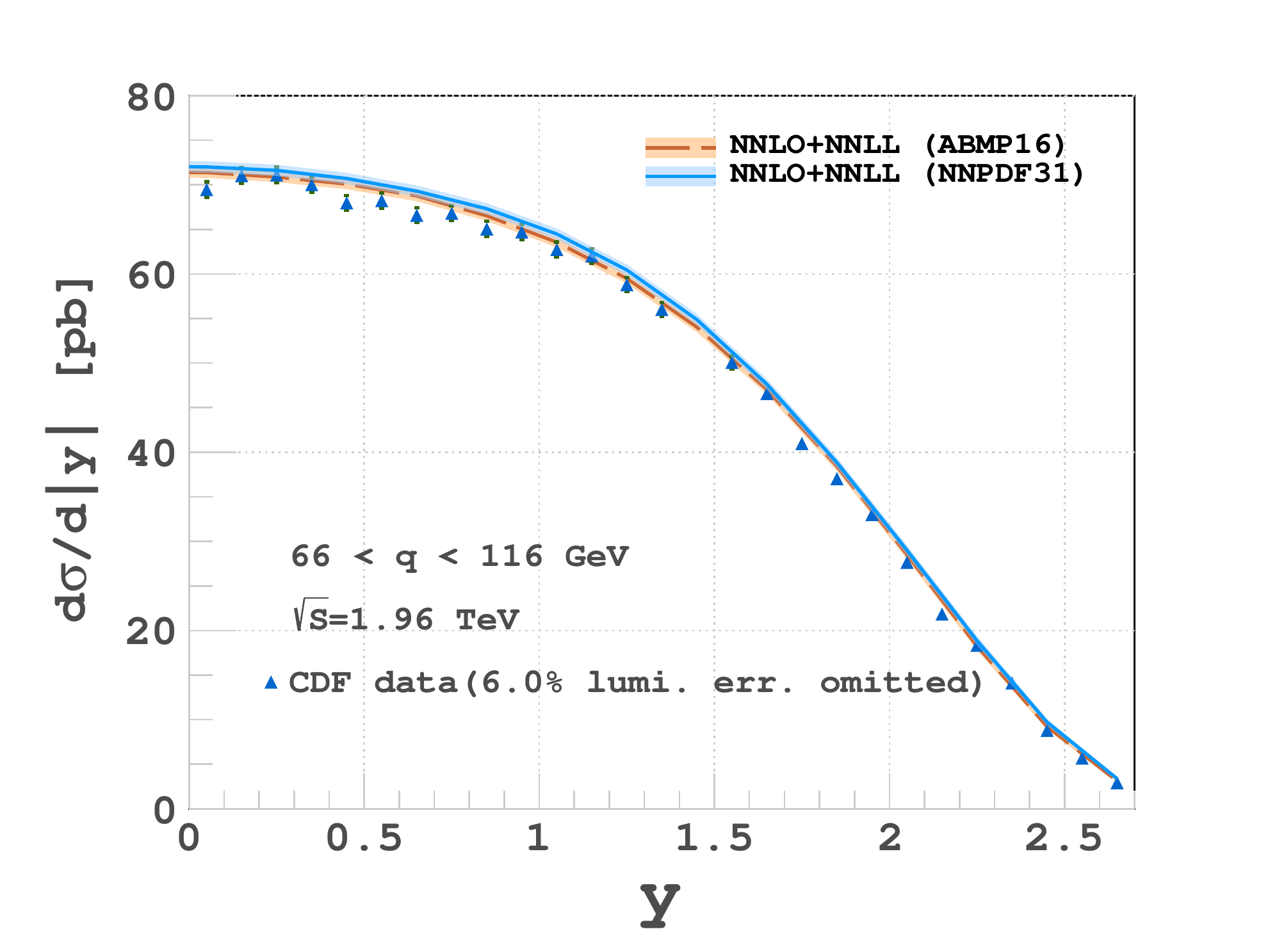}
\caption{Comparison between the resummed results and the CDF data~\cite{Affolder:2000rx,Aaltonen:2010zza} at $\sqrt{s}$=\{1.8 TeV, 1.96 TeV\} in the invariant mass range  $66<q<116$ GeV for two different PDF sets.
}\label{fig:CDF}
\end{figure}
\end{center}
\end{widetext}

Both at Tevatron and at the LHC, there are already precise measurements of rapidity 
distributions for different ranges of invariant mass $q$. For one of the earliest set of measurements see
NuSea~\cite{Towell:2001nh,Webb:2003bj}. Since the data from the  LHC depends heavily on the kinematic cuts 
of the final states we cannot directly compare against our predictions. On the other hand
CDF \cite{Affolder:2000rx,Aaltonen:2010zza} has data for the rapidity distributions for wide range of $y$ with 
invariant mass range $66<q<116$ GeV. In Fig.~\ref{fig:CDF}, we have compared our predictions against the
data at $\sqrt{s}=1.8$ TeV and at $\sqrt{s}=1.96$ TeV after integrating $q$ between the above mentioned range for two different choices 
of PDF sets. The scale uncertainty is obtained as before by using 7-point scale variations around the central 
value $\mu_R=\mu_F=M_Z$.  We note that at NNLO+NNLL level, the resummed contributions over the fixed order
is very mild, less than 0.5\%. We have also observed that the resummed effects
become significant for large invariant mass regions.

\textit{Discussion and Conclusion}.--- 
In this article we have done a detailed study on the role of resummed threshold logarithms  
for the rapidity distribution of pairs of leptons in the Drell-Yan process at the LHC.
Being one of the cleanest channels at the hadron colliders, precise
measurements of various observables such as inclusive cross section, 
transverse momentum and rapidity distributions are already available.
Precise predictions from perturbative QCD are known to NNLO level for long and corrections
from electroweak theory have become available in recent times.  The latter effects being
close to the second order effects from QCD so that dedicated efforts to understand the EW
effects have been undertaken.  Owing to the dominant QCD interactions, soft gluons play
vital role in most of the observables.  They show up in certain kinematic regions through
large logarithms in the perturbative computations.  Often they spoil the reliability of
the fixed order predictions.  In this present article, we have made a detail study on the effect of
these soft gluons within the resummation framework.  In the literature 
two different approaches exist.  They differ in the kind of logarithms that are resummed to
all orders.  The approach which uses Mellin-Fourier transformation to achieve the resummation to
resum large logarithms of the scaling variable $z$ has been well studied for the rapidity
distribution.  Threshold logarithms resulting from regions where the scaling variable $z_i$ approach
unity are successfully resummed using M-M approach.  We used the latter approach
to get the quantitative predictions at NNLO+NNLL level.  Since these formalisms resum different
type of logarithms to all orders, they are expected to give different numerical predictions.    
In this article, we have not only undertaken a detailed study on the numerical impact of
M-M approach for the first time for the Drell-Yan process but 
also made a detailed numerical comparison against
the M-F approach.  While at LL and NLL level, they differ very much, surprisingly at
the NNLL level both the approaches converge to a few percent correction to the fixed order prediction.
This could be accidental, however it is desirable to understand this coincidence at NNLL level. 
Our numerical study on the dependence of renormalisation and factorisation scales shows that
the optimal central scales for the resummed  result are $\mu_R=M_Z/2$ and $\mu_F=M_Z$ while
it is $\mu_R=\mu_F=M_Z$ for NNLO.  We have also found that, for wide range of rapidity, the scale
uncertainties from NNLL contributions at every order are slightly larger than those from fixed order
results.  We believe that this could be due to an incomplete cancellation of  scale dependent terms between
resummed result and the PDFs.  Note that the PDFs that we use are extracted from data using the
fixed order perturbative predictions for the observables and also using evolutions equations controlled
by splittings functions computed to desired order in strong coupling constant.  Hence, we expect
that there will be a better cancellation of scale if appropriate resummed PDF sets are available.  
We have also presented our predictions for various choices of PDFs from various PDF groups.  
Each group has several sets and hence we have not only made comparisons with respect to
various groups but also estimated the uncertainty from different sets within each PDF group.
We have also predicted the q integrated rapidity distribution.  Since our resummation
formalism can not take into the experimental cuts such as  the transverse momentum and/or polar angles
of the final state leptons, we can not make any direct comparisons with the existing data on the q integrated
rapidity distribution measured at the LHC which are extracted after 
employing cuts on transverse momentum of final state
leptons.  On the other hand we have compared our predictions against CDF data at Tevatron for the invariant mass range $66<q<116$ GeV and found good agreement within both theoretical and experimental uncertainties. We believe that perturbative results that take into account both fixed order as well as the
resummed contributions will provide a precise determination of PDFs from the ample data that are already
available at the LHC. 

\textit {Acknowledgement}.--- 
We thank S. Alekhin, J. Bl\"umlein, S. Moch and F.\ J.\ Tackmann for suggestions and careful reading 
of the manuscript. 
We are thankful to S. Catani, M. Grazzini, M. Neubert, G. Ferrera,  A. Vicini, M. Bonvini and
 L. Rotolli, W. Vogelsang for useful discussions. GD would like to thank J. Michel and L. Dixon  for fruitful 
 discussions. PKD would like to thank G. Ridolfi, R. Harlander and F. Maltoni for useful
 suggestions to improve the presentation of the paper. PB would like to thank T. Becher and
L. Magnea for useful discussions. Ravindran would like to thank M. Neubert for his 
kind hospitality at University of Mainz where  part of the work was carried out.
We also thank P. Mangalapandi for his help related to cluster computing system at IMSc. 
GD acknowledges research support from DESY.

\bibliography{dyres}

\bibliographystyle{apsrev4-1}
\end{document}